\documentclass{mn2e}
\usepackage{epsfig}

\def\ngc{{NGC 4051}}

\def\xmm{{\it XMM-Newton}}
\def\chandra{{\it Chandra}}

\def\et{{et al.\ }}

\def\suzaku{{\it Suzaku}}

\newcommand{\ls}{\mathrel{\hbox{\rlap{\hbox{\lower4pt\hbox{$\sim$}}}\hbox{$<$}}}}
\newcommand{\gs}{\mathrel{\hbox{\rlap{\hbox{\lower4pt\hbox{$\sim$}}}\hbox{$>$}}}}

\def\H0{{\rm ~km~s^{-1}~Mpc^{-1}}}

\def\et{{et al.}}

\def\deg{^\circ}

\title[X-ray spectrum of \ngc]
{An extended \xmm\ observation of the Seyfert galaxy \ngc\ 
- III. FeK emission and absorption}
\author[K.A.Pounds \et]
        {K.A.Pounds,
	S.Vaughan 
	\\
Department of Physics and Astronomy, University of Leicester,
Leicester, LE1 7RH, UK\\}

\date{Accepted ; Submitted }
\pagerange{\pageref{firstpage}--\pageref{lastpage}}
\pubyear{2010}
\begin{document}
\maketitle
\label{firstpage}

\begin{abstract}
An extended \xmm\ observation of the Seyfert 1 galaxy \ngc\ in 2009 detected a photoionised outflow with a complex absorption line velocity structure and
a broad  correlation of velocity with ionisation parameter, shown in Pounds \et (2011) to be consistent with a highly ionised, high velocity wind running
into the interstellar medium or previous ejecta, losing much of its kinetic energy in the resultant strong  shock. In the present paper we examine the FeK
spectral region in more detail and find support for two distinct velocity components in the highly ionised absorber, with values corresponding to the
putative fast wind ($\sim$0.12c) and the post-shock flow (v$\sim$5000-7000 km s$^{-1}$). The Fe K absorption line structure is seen to vary on a
orbit-to-orbit timescale, apparently responding to both a short term increase in ionising flux and - perhaps more generally - to changes in the soft X-ray
(and simultaneous UV) luminosity. The latter result is particularly interesting in providing independent support for the existence of shocked gas being
cooled primarily by Compton scattering of accretion disc photons. The Fe K emission is represented by a narrow fluorescent line from near-neutral matter,
with a weak red wing modelled here by a relativistic diskline. The narrow line flux is quasi-constant throughout the 45-day 2009 campaign, but is
resolved, with a velocity width consistent with scattering from a component of the post-shock flow. Evidence for a P Cygni profile is seen in several
individual orbit spectra  for resonance transitions in both Fe XXV and Fe XXVI.  
\end{abstract}


\section{Introduction}

An extended \xmm\ observation of the Seyfert 1 galaxy \ngc\ in 2009 found a rich absorption line spectrum indicating a photoionised outflow with a wide
range of velocities and ionisation parameter (Pounds \et\ 2011; hereafter Paper I).  The absorption line velocity structure and a broad  correlation of
velocity with ionisation parameter were shown to be consistent with an outflow scenario where a  highly ionised, high velocity wind, perhaps launched
during intermittent super-Eddington accretion (King and Pounds 2003), runs into the interstellar medium or previous ejecta, losing much of its kinetic
energy in the resultant strong  shock. 

The wider importance of such shocked outflows lies in the possibility that the accumulated thrust from multiple episodes - rather than the outflow energy
- would eventually drive gas from the bulge, thereby limiting further star formation and black hole growth. Such a momentum-driven feedback mechanism has
been shown by King (2003, 2005) to reproduce the observed  correlation of black hole and bulge mass (e.g. Ferrarese and Merritt 2000, Gebhardt \et\ 2000,
Haring and Rix 2004)). 

Growing evidence for extreme velocity (v$\sim$0.1-0.2c) X-ray outflows (Chartas \et\ 2002, Pounds \et\  2003, 2006; Reeves \et\ 2003, Cappi 2006, Tombesi
\et\ 2010) has been confined to the very highly ionised matter (log$\xi$$\sim$3.5--4) most readily detected in the Fe K band.  If the Tombesi \et\
findings are confirmed, with highly ionised outflows at v$\sim$0.1c being relatively common in bright nearby galaxies, then perhaps Eddington or mildly
super-Eddington accretion is also more common than generally believed. (A case for AGN black hole masses being over-estimated has recently been argued by
King (2010a)). 

We note that if the ejection of fast outflows is intermittent, as suggested in Paper I for \ngc, then only where such a  wind is current or was launched
very recently will it retain a line-of-sight column close to the value  N$_{H}$$\sim$$10^{24}$ cm$^{-2}$, appropriate to continuum driving (King and
Pounds 2003) and probably required for detection of a v$\sim$0.1c flow in a very low redshift AGN with current X-ray missions.

The primary aim of the present paper is to examine the complex Fe K emission and absorption profile of \ngc\ and attempt to resolve and identify dicrete spectral features
carrying information on the highly ionised outflow and re-processing of the primary X-ray continuum. We have sought to confirm the presence of the highly
ionised pre-shock wind discussed in Paper I, for which scaling from the strong absorption seen at velocities up to $\sim$9000 km s$^{-1}$ 
predicts a wind velocity in the range $\sim$0.11--0.13c. We also examine the Fe K absorption line structure at different continuum flux levels to
establish whether the highly ionised outflow component varies on a similar few-day timescale to that observed in RGS spectra for lower ionisation gas over the
same velocity regime (Pounds and Vaughan 2011a; hereafter Paper II). 

\section{Observations and data analysis}

\ngc\ was observed by \xmm\ on 15 orbits between 2009 May 3 and June 15. Here we use primarily data from the EPIC pn camera (Str\"{u}der \et 2001),  which
has the best sensitivity of any current instrument in the Fe K band ($\sim$6-10 keV). Excluding high background data near the end of each orbit the total
exposure  available for spectral fitting is $\sim$600 ks.  Further details on the timing, flux levels and X-ray light curves for each orbital revolution
are included in Vaughan \et\ (2011). 
Line energies are quoted throughout the paper adjusted for the redshift of \ngc, with the calculation of line shifts
and velocities also taking account of a pn detector energy offset (see below) by cross-reference to the simultaneous data from the MOS camera (Turner \et
2001). Line widths are the intrinsic values, after allowance for pn camera energy resolution. Unless specified otherwise,  uncertainties correspond to 90
$\%$ confidence limits.

\subsection{Spectral structure in the Fe K region}

Figure 1 (top panel) shows EPIC pn spectral data summed over the whole 2009 \xmm\ observation, plotted as a ratio to the continuum determined by first
fitting the data between 5 and 10 keV with a power law plus Gaussian emission line at $\sim$6.4 keV, and then removing the line. The aim of that simple
procedure was to delineate any emission and absorption fine structure without introducing potential artefacts from modelling the characteristic curvature
in the continuum  with reflection or ionised absorption. 

The ratio plot of pn data to this simple power law continuum shows a narrow Fe K emission line with an apparent red wing, together with several absorption
lines to higher energy. Gaussian fitting to the individual spectral features finds the Fe K emission can be modelled with a narrow and a broad component.
The narrow line energy (adjusted for the redshift of \ngc) is 6.437$\pm$0.006 keV (\footnote{The summed MOS spectrum finds a narrow line energy of
6.409$\pm$0.004 keV, consistent with fluorescence from stationary and cold matter. The MOS energy calibration is believed to be more stable than that of
the pn and all absorption line blue-shifts used here to calculate outflow velocities are adjusted by a corresponding factor of 0.996}). The broad
component is centred at 6.4$\pm$0.1 keV with width $\sigma$=650$\pm$70 eV. 

Three absorption  lines, at $\sim$6.8, $\sim$7.15 and $\sim$8.0 keV and a possible fourth line at $\sim$7.75 keV were  then modelled with negative
Gaussians. Lines  labelled 1,3 and 4 in figure 1 were found to be narrow and were therefore fixed at a width corresponding to the pn resolution
($\sigma$$\sim$60 eV).  Component 2 is clearly resolved, and the Gaussian fit yielded an intrinsic 1$\sigma$ width of 113$\pm$17 eV. Measured line
energies (adjusted to the rest  frame of \ngc) are listed in Table 1 where alternative identifications are used to infer relevant outflow velocities.

\begin{figure}                                                          
\centering                                                              
\includegraphics[width=6.28cm, angle=270]{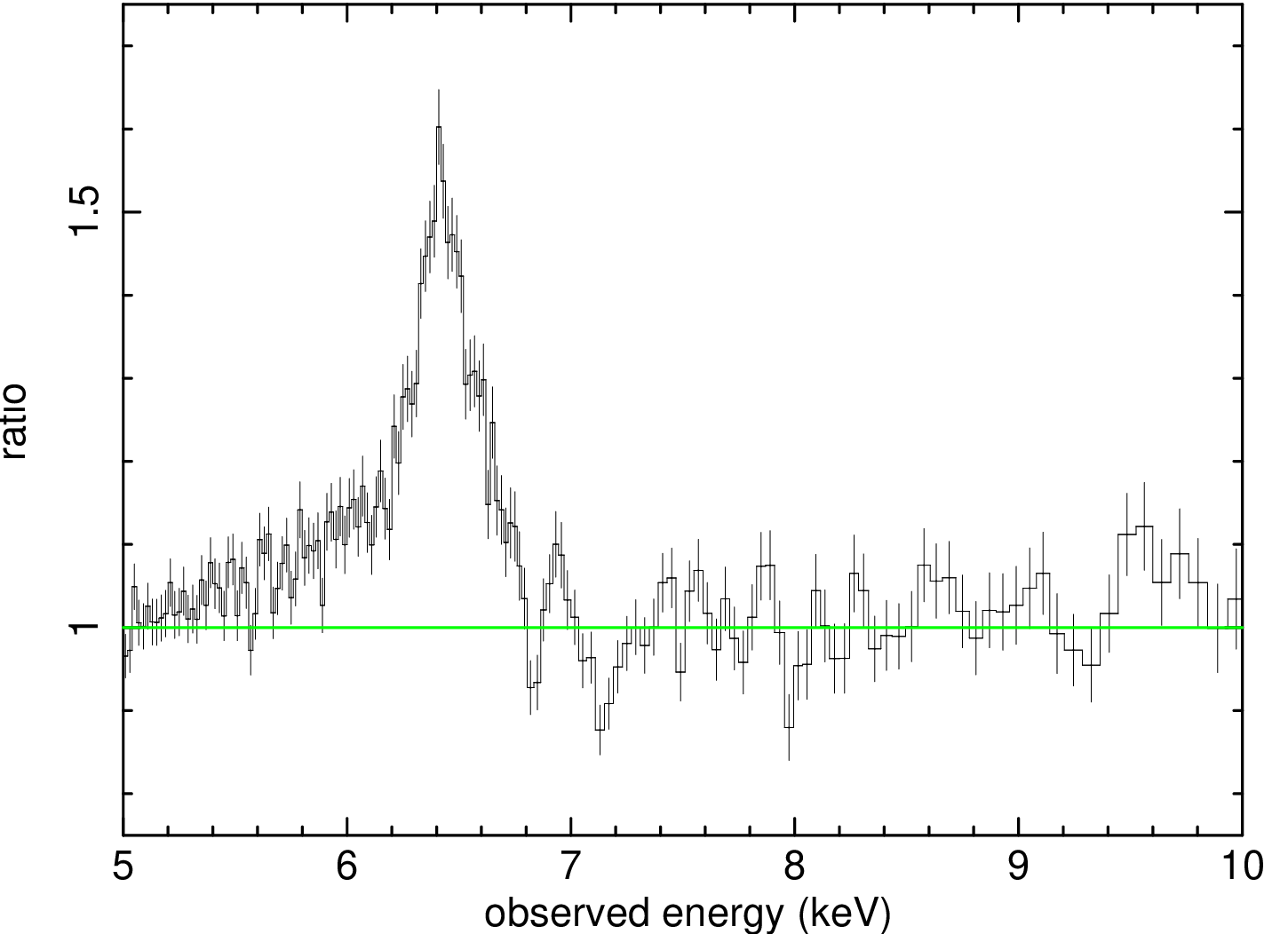}                                            
\centering                                                              
\includegraphics[width=6.28cm, angle=270]{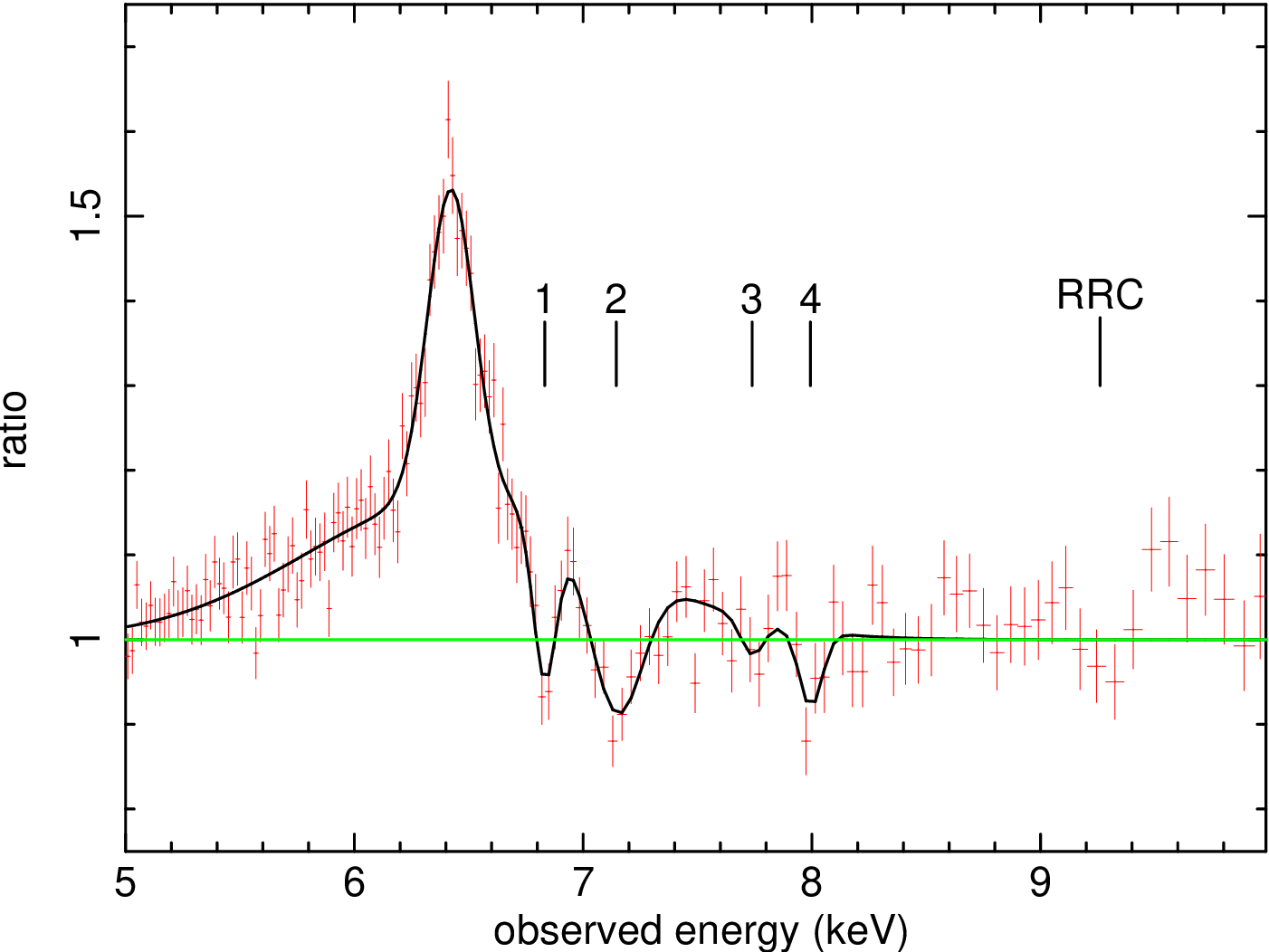}                                                                                        
\caption                                                                 
{pn data summed over the whole 2009 observation plotted as a ratio to a simple power law between 5 and 10 keV (details in the text).  Gaussian fitting in
the lower panel models the Fe K emission profile and several possible absorption lines. Details of  the Gaussian fits and possible identifications of the
absorption lines are listed in Table 1. The zero velocity threshold energy of the putative Fe XXVI RRC is also marked.}      
\end{figure}

Component 1 is rather unambiguously identified with the resonance line of He-like Fe XXV (although a more blue-shifted inner shell UTA is theoretically
possible, Kallman \et\ 2004). The Fe XXV identification yields a projected outflow velocity of $\sim$5500 km s$^{-1}$, similar  to the  strong absorption
seen in O VIII and higher level ions in the RGS data, and identified with the early post-shock flow in Paper I.

Component 2 has two candidate identifications, the more likely being with H-like Fe XXVI, corresponding to an outflow velocity of $\sim$7200 km s$^{-1}$ 
for the measured centroid energy. However, as the line profile is broadened (as it also was in the \xmm\ observation of 2001 (Pounds \et\ 2004)), an alternative 
fit to component 2 is a velocity range  $\sim$5000--9000 km
s$^{-1}$ for Fe XXVI Lyman-$\alpha$. Interestingly, a similar broad absorption/velocity trough is seen in O VIII Lyman-$\alpha$ in the highest flux RGS spectrum
(Paper I). 

The third strongest absorption line in figure 3 is component 4, at $\sim$8 keV. The most interesting identification, in the context of the present study,  is with Fe
XXVI  Lyman-$\alpha$, yielding an outflow velocity of 38300$\pm$1250 km s$^{-1}$, within the range predicted in Paper I for the pre-shock wind. However,
lower velocity absorption from the FeXXV 1s-3p transition must be present at some level, though the relative line strength and low velocity, compared with the 1s-2p line
at $\sim$6.8 keV, suggest this may not be the dominant  absorption component. Of the remaining candidates listed in Table 1, the Fe XXV 1s-2p line appears
least likely, implying an extreme velocity, well outside the predicted range for the putative pre-shock wind, while significant absorption from
He-like Ni would imply a substantial over-abundance of Ni  relative to Fe. 

Finally, if confirmed, component 3 would provide some independent support for the detection of the pre-shock wind, the most likely 
identification with the Fe XXV
resonance line then forming a high velocity pairing with that of Fe Lyman-$\alpha$ for component 4.   

The ratio plot of figure 1 also shows a positive feature at $\sim$9.4 keV lying close to the ionisation threshold energy of Fe XXVI.   An intriguing
identification, given the strong radiative recombination continua (RRC) seen in the  RGS spectrum (Paper II), would be with the
RRC of Fe XXVI, with a blue-shifted velocity of $\sim$5300$\pm$2000 km s$^{-1}$. However, we see in the next Section
that the reality of this emission feature is dependent on correctly modelling the underlying continuum near 9 keV.

In summary, a simple examination of pn spectral data summed over the $\sim$600 ks observation of \ngc\ finds evidence for several blue-shifted absorption
lines in the Fe K band, together with both narrow and broad emission components. The  implied outflow velocities from absorption lines 1 and 2, and the
indication of line broadening in the feature at $\sim$7.1 keV, maps closely to the velocity profile observed in OVIII Lyman-$\alpha$ and similar lower
mass ions (Paper I), suggesting we are seeing in Fe K a co-moving, more  highly ionised (lower density)  component of the same  post-shock flow. 
Of particular interest in the context of the shocked wind interpretation of the complex RGS absorption spectrum (Paper I, Paper II) is the possible
identification of absorption line 4 with Fe XXVI Lyman-$\alpha$, corresponding to a velocity component with  the predicted (factor $\sim$4) higher value
for the putative pre-shock wind. We note, however, that outcome must be qualified by a likely blend with the 1s-3p transition in Fe XXV.

\begin{table}
\centering
\caption{Summary of Gaussian line fits to the four most significant absorption lines in figure 1, adjusted to the \ngc\ rest frame, and with the
corresponding outflow velocity also allowing for the known pn calibration error}
\begin{tabular}{@{}lccc@{}}
\hline
line & energy (keV) & ident  & velocity (km s$^{-1}$ \\
\hline
1 & 6.85$\pm$0.01 & FeXXV (1s-2p) &  5500$\pm$450 \\
2 & 7.16$\pm$0.015 & FeXXV (1s-2p) &  17900$\pm$650 \\
2 & 7.16$\pm$0.015 & FeXXVI Ly-$\alpha$ &  7200$\pm$650 \\
3 & 7.76$\pm$0.045 & FeXXV (1s-2p) &  40000$\pm$1750 \\
3 & 7.76$\pm$0.045 & FeXXVI Ly-$\alpha$ &  29800$\pm$1750 \\
4 & 8.01$\pm$0.03 & FeXXV (1s-2p) &  48200$\pm$1250 \\
4 & 8.01$\pm$0.03 & NiXXVII (1s-2p) &  8300$\pm$1250 \\
4 & 8.01$\pm$0.03 & FeXXV (1s-3p) &  3700$\pm$1250 \\
4 & 8.01$\pm$0.03 & FeXXVI Ly-$\alpha$ & 38300$\pm$1250 \\
\hline
\end{tabular}
\end{table}

\section{Fitting with Xspec}

The above procedure, of examining spectral structure in the Fe K band against a simple power law baseline, was adopted to circumvent possible confusion
with  discrete spectral features that might be introduced in seeking to model the observed curvature in the \ngc\ broad band spectrum below  $\sim$6 keV
and the step function decrease seen above $\sim$7 keV.  

To assess such effects and further quantify the emission and absorption line structure in the Fe K spectrum, we have repeated the Fe K profile analysis, but now
modelling the underlying continuum with a power law, part of which is attenuated by an ionised absorber, together with a low ionisation reflection
component (required for consistency with the narrow emission line at $\sim$6.4 keV). To again avoid the introduction of fine structure into the continuum
fit, we model continuum reflection with {\it pexriv} (Magdziarz and Zdziarski 1995) and ionised absorption with {\it absori} (Done \et\ 1992), and use the
{\it Xspec} software (Arnaud \et\ 1996) to seek an acceptable continuum fit to the EPIC data over the 3-10 keV band. We term this the
{\it pcref} model.

Initial spectral fitting at different flux levels showed the flux of the narrow Fe K line to remain essentially unchanged throughout the 2009 observation,
suggesting that a neutral or weakly ionised continuum reflection component should also stay constant.  Fitting the 3-10 keV pn
spectrum of the lowest flux orbit, rev 1739, with the {\it pcref} model then provided a measure of that quasi constant reflection continuum. The spectral
fit for rev 1739 is reproduced in figure 2, along with the very similar fit for the extended low flux observation in 2002. The {\it pexriv} parameters
from rev 1739 were then carried forward as a description of the weakly ionised reflection component in fitting the continuum for all 2009  spectra.  We
focus initially on modelling the summed 2009 pn data, to allow direct comparison with the FeK spectral features obtained in the previous Section. 

\begin{figure}                                                          
\centering                                                              
\includegraphics[width=6.28cm, angle=270]{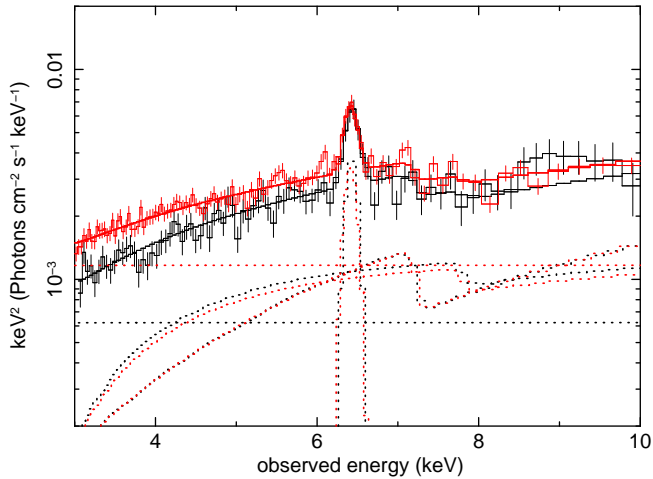}                                                                                                                                                                                                                                                                        
\caption                                                                 
{(top) Unfolded spectral fits to the low flux data of rev 1739 in 2009 (black) and rev 0541 in 2002 (red) used here to parameterise the quasi-constant 
reflection components in the \ngc\ spectrum}      
\end{figure} 

In modelling the summed pn data the power law index was tied for both absorbed and unabsorbed direct components and for the input to {\it pexriv}.
Abundances were fixed at solar values and the redshift of the reflected continuum set at that of \ngc, as in the fit to the low  flux data. The common
power law index was left free, as were the normalisations of both direct continuum components, and the column density, ionisation parameter and `redshift'
(a proxy for the velocity) of the partial covering absorber.  

The parameters of the continuum best fit include a tied power law index $\Gamma$=1.99$\pm$0.01, close to the value found for the dominant component in a
study of broad-band spectral variability (Vaughan \et\ in preparation). The continuum reflection, carried over from the low flux fit, is now less dominant
in comparison to the direct continuum components, but still contributes significantly to the spectral curvature and provides the step function change seen
in the raw data at $\sim$7.1 keV. An ionised gas column of  N$_{H}$=12$\pm$2$\times 10^{22}$ cm$^{-2}$ covers $\sim$50 \% of the direct continuum, with
blue shift of 0.026$\pm$0.009 and ionisation parameter $\xi$=210$\pm$17 erg cm s$^{-1}$ combining to place a (weak) absorption edge near $\sim$8 keV.

With the addition of a positive Gaussian  to model the narrow Fe K emission line near 6.4 keV, the 3-10 keV spectral fit remained unacceptable,  with
$\chi^{2}$ of 1280 for 1017 degrees of freedom. The main residuals are shown in figure 3 (top panel), where the narrow emission line has been
re-introduced to allow visual  comparison with figure 1. Inclusion of the cold reflection continuum now confines the broad excess emission to the red side
of the narrow Fe K line.  As the latter is strongly suggestive of a relativistic fluorescent emission component, we then modelled this excess with a {\it
diskline} (Fabian \et\ 1989), improving the 3--10 keV fit to $\chi^{2}$ of 1147/1012. We term this the {\it pcrefdisk} model and review the emission
components in Section 4.

To quantify the residual Fe K spectral structure (figure 3, lower panel) the {\it pcrefdisk}
model was  then taken as a baseline, with positive and negative Gaussians  added in {\it Xspec} to model the emission and absorption fine
structure. All parameters of the continuum fit (other than for {\it pexriv}) were left free in order to quantify the improvement in fit with the addition
of each discrete spectral component.   The final outcome was an excellent 3-10 keV spectral fit, with $\chi^{2}$ of 994 for 1000 degrees of freedom. The
unfolded spectrum and spectral components are illustrated in figure 4.  (\footnote{To independently test the reliability of detection of absorption lines we used a
$\Delta \chi^2 > 11$ criterion, the improvement in the fit expected to occur with a probability $p = 0.01$  (or `$99$ per cent confidence') when adding a
narrow absorption line in the range $6.7-9$ keV, when no such line is present. The $p$-value was calibrated using a Monte Carlo method. We simulated
$1000$ EPIC pn source and background spectra with the same exposure time as the summed 2009 spectrum, the spectra being generated using a simple model
comprising a power law plus narrow, neutral iron line, with parameters derived from fitting the real data, but randomised using the covariance matrix at
the best fit. (The technical details of that procedure are discussed in Hurkett et al. 2008; section 3.2). For each simulated spectrum we fitted the $5-10$ keV ranges with
the power law model plus emission line, and then inserted a narrow absorption line with centroid energy restricted to the $6.7-9$ keV band. For each
simulation we recorded the $\min(\chi^2)$ statistic for the two models (line and no line), from which we estimated the sampling distribution of the
$\Delta \chi^2$ statistic. A $\Delta \chi^2$ of $>11$ occurred for only $8$ of $1000$ simulations.})

\begin{figure}                                                          
\centering                                                              
\includegraphics[width=6.28cm, angle=270]{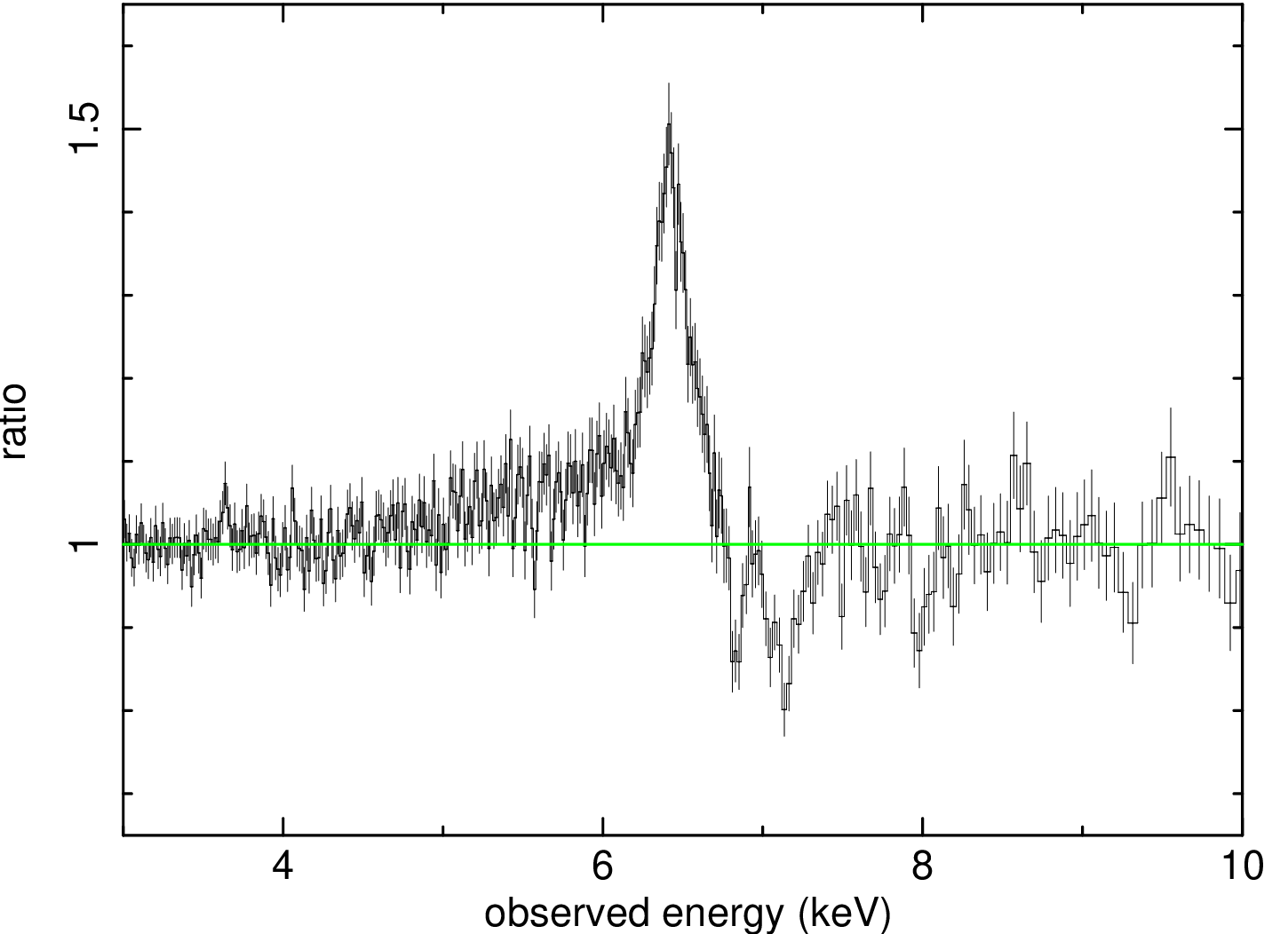}                                                                                                                                    
\centering                                                              
\includegraphics[width=6.28cm, angle=270]{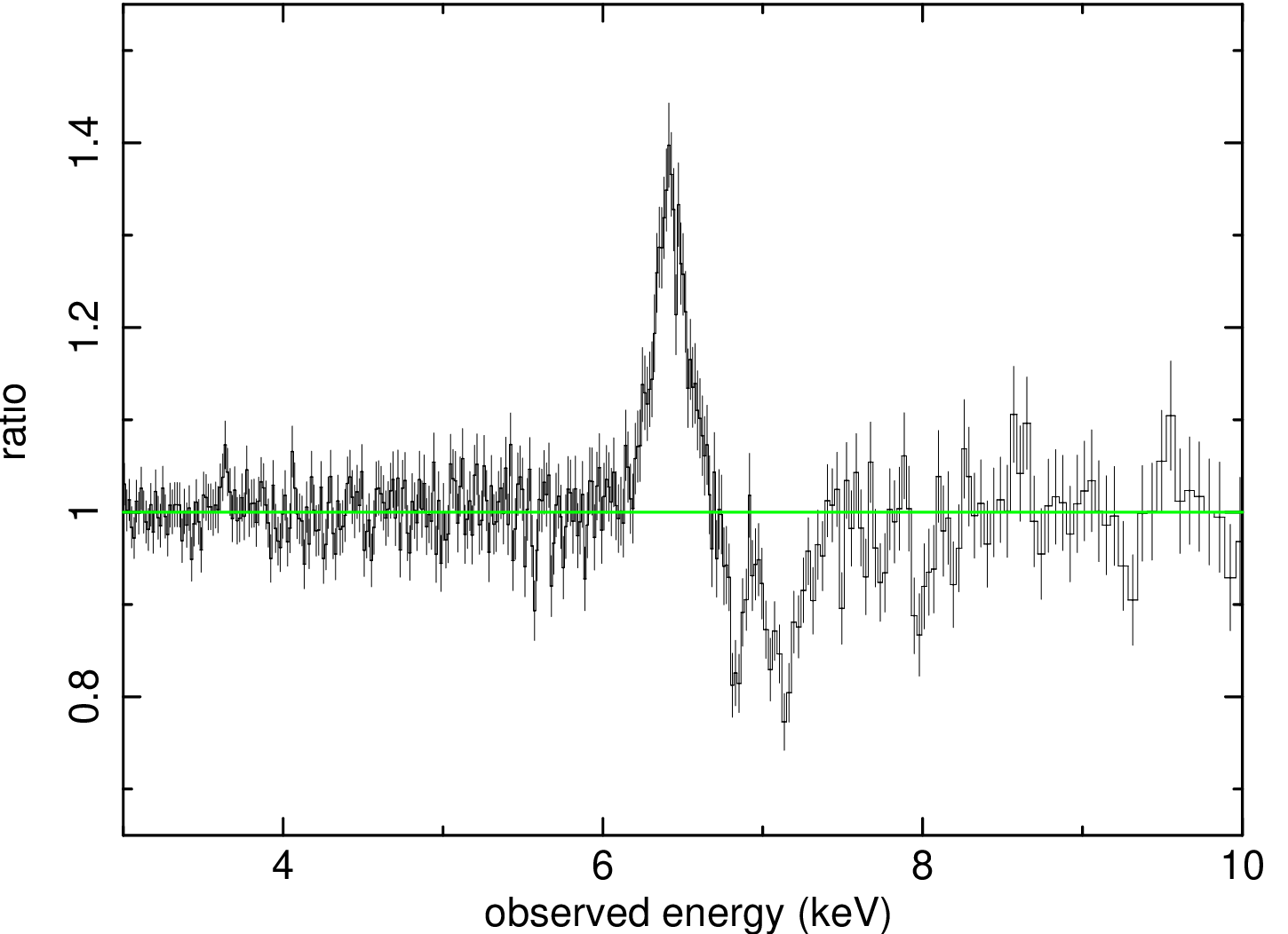}                                                                                                                                    
\caption                                                                 
{(top) Data residuals for the {\it pcref} continuum model of the summed 2009 pn spectrum, and (lower) with the addition of a relativistic emission line. 
Details in the text}       
\end{figure} 

The absorption line energy, equivalent width and significance of each absorption line are listed in Table 2. Each line is again unresolved apart from that
near 7.15 keV.   While that line broadening might be due to imperfect modelling of the {\it pexriv} absorption edge, a more interesting alternative is that
noted in the previous Section, namely that the line width relates to a  spread of outflow velocities, as found in the same post-shock velocity regime in
RGS data (Paper I).

In summary, a more physical spectral analysis finds aborption lines 1, 2 and 4 to be detected at high significance, with line 3 at $\sim$2$\sigma$.  On
the critical question of the identification of line 4, support for detection of the putative pre-shock wind is strengthened  by the better agreement
between the derived (high) velocities for components 3 and 4, while the EW ratio of 1.5$\pm$0.5 for components 1 and 4 is inconsistent with the (optically
thin) ratio of 5.2 for FeXXV alone. The velocities obtained from components 1 and 2 remain marginally inconsistent; however, that difference may be real,
if the  highly ionised post-shock flow component has a velocity/ ionisation gradient similar to that found in the RGS data (Paper I).

\begin{figure}                                                           
\centering                                                              
\includegraphics[width=6.28cm,angle=270]{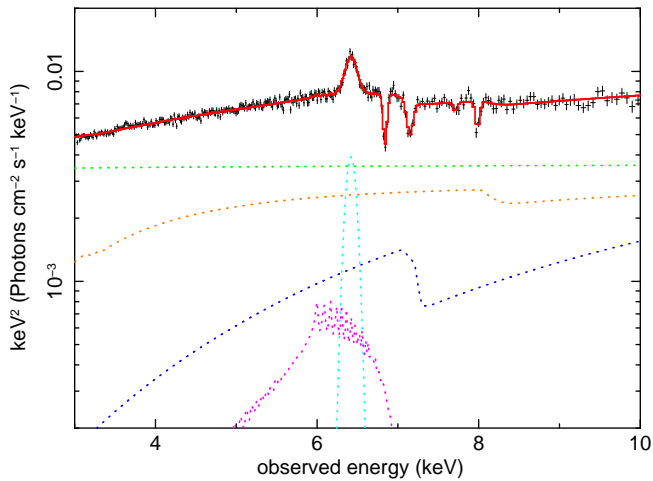} 
\caption  
{Unfolded spectrum from the broad band fit to the pn data summed over the whole 2009 observation, with the spectral curvature and step function drop in
the continuum above 7 keV modelled by a constant reflection component carried forward from the low flux fitting, together with a direct power law
component  partially covered by a photoionised absorber. The Fe K emission is modelled by a narrow Gaussian line and a relativistic
{\it diskline}. Several absorption lines are required to obtain a good overall fit to the data}       
\end{figure}

\begin{table}
\centering
\caption
{Summary of Xspec modelling of the absorption line spectral structure shown in figure 3. Line energies are adjusted only for the redshift of
\ngc, with the corresponding outflow velocity (assuming the preferred identifictions with 1s-2p and 1s-3p transitions of Fe XXV and with Fe XXVI
Lyman-$\alpha$) also allowing for the known pn calibration error. Line equivalent widths are determined against the direct continuum, excluding the cold
reflector}
\begin{tabular}{@{}lcccc@{}}
\hline
line & energy(keV) & velocity (km s$^{-1}$ & EW(eV) & $\Delta$$\chi^{2}$  \\
\hline
1 & 6.86$\pm$0.01 & 5500$\pm$500 & -34$\pm$5 & 43 \\
2 & 7.17$\pm$0.03 & 7300$\pm$1600 & -50$\pm$5 & 80 \\
3 & 7.74$\pm$0.08 & 39100$\pm$3200 & -11$\pm$5  & 7 \\
4 & 8.00$\pm$0.05 & 3400$\pm$1500 & -23$\pm$5 &  19 \\
4 & 8.00$\pm$0.05 & 37900$\pm$1500 & -23$\pm$5  & 19 \\
\hline
\end{tabular}
\end{table}

\section{Fe K emission}

The broad band spectral fitting in the previous section modelled the Fe K emission with a narrow  Gaussian component and a relativistic {\it diskline}
(Fabian \et\ 1989). The narrow component (when adjusted for the pn calibration shift) lies close to 6.4 keV, with an equivalent width (EW) against the direct
continuum of $\sim$110 eV, consistent with fluorescence from the weakly ionised  reflector in the continuum fit (Matt 2002).  Finding parameters of the
narrow emission line in the summed data consistent with those for the low flux orbit, rev 1739, is strong  support for  our initial assumption that the
cold reflection component was essentially unchanged during the 2009 campaign. However, the line appears resolved, with an intrinsic width
1$\sigma$=70$\pm$10 eV. If due only to velocity broadening, that width corresponds to $\sim$7500$\pm$1100 km s$^{-1}$ (FWHM), intriguingly close to the
dominant velocity in the post-shock outflow.

Relativistic fluorescence emission from the inner accretion disc is a reasonable candidate for the red wing of the Fe line, given that the rapid flux
variability in the X-ray  emission from   \ngc\ (Vaughan \et\ 2011) indicates a primary continuum source dimension of a few 10R$_{G}$.  The 3-10 keV
spectral fit described in the previous Section includes a {\it diskline} with a input line energy of 6.8$\pm$0.2 keV, emissivity index
$\beta$=-4.5$\pm$0.7, R$_{in}$=6$\pm$1 and  inclination of 23$\pm$3 $\deg$. The line is relatively weak, with EW $\sim$130 eV.  

Although the narrow emission line appears quasi-constant throughout the 2009 observation, we cannot assume that is also true for the {\it diskline} 
component. That issue is raised in the following Section when we explore the variability of the Fe K profile and find evidence for  PCygni emission
associated with  the stronger absorption lines. If confirmed, that would confirm a substantial covering factor in the ionised flow (as was found
previously for the high velocity outflow in the luminous Seyfert PG1211+143; Pounds and Reeves, 2009).

\section{Short term variability in the Fe K profile}

The Fe K absorption line spectrum in the summed 2009 data is more complex than the single $\sim$7.1 keV absorption line reported from the 2001 \xmm\
observation (Pounds \et\ 2004). In particular, a second strong absorption line at $\sim$6.8 keV supports the 2001 line identification with Fe XXVI
Lyman-$\alpha$, for a velocity of $\sim$6500 km s$^{-1}$ (in preference to Fe XXV 1s-2p and a velocity of $\sim$17500 km s$^{-1}$), with  a similar
outflow velocity but somewhat lower ionisation parameter in the summed data from 2009. A coordinated \chandra\ and \suzaku\ study 6 months before our
2009 \xmm\ campaign also found absorption lines at $\sim$6.8 and $\sim$7.1 keV (Lobban \et\ 2010), perhaps suggesting little  change in the
highly ionised $\sim$6500 km s$^{-1}$ outflow component over that timescale.

However, our analysis of the RGS data from the 2009 \xmm\ observation (Paper I, Paper II) found evidence for a change in the ionisation state 
of more moderately ionised gas on a timescale of a few days. As the two strongest absorption lines in the Fe K region indicate a comparable
($\sim$5000-7000 km s$^{-1}$) velocity
regime, it is interesting to check for a similarly  rapid response  to changes in the incident continuum flux level. More particularly, if the Fe XXV and
XXVI absorption arises in a co-moving, lower density (and more highly ionised) flow component, we might anticipate measurable variability over the 6-weeks
duration of the 2009 observation.  

To search for short-term variability in the Fe K profile we have used the summed 2009 data fit described in Section 3 as a template against which to compare
spectral fits for individual orbits. The cold reflection and narrow Fe K emission line parameters were fixed in this comparison, but the  power law index
and normalisations, {\it diskline} normalisation and ionisation parameter of the partial cover were left free. The energy and width of the individual
absorption lines were fixed at the summed data values, implying no short-term change in the velocity structure of the outflow.  Fitting was
carried out for individual orbit data over 3-10
keV.

Figure 5 plots the equivalent width (EW) of the 4 absorption lines of figure 4 and Table 2, together with the 0.3-3 and 7-10 keV luminosities, for
successive orbits rev1721 (1) to rev 1743 (15), with the summed spectral fit as (16).  The single orbit parameters were then compared to the best fitting
constant value, using a $\chi^{2}$ test in order to quantify evidence for their variability. In three cases ($\sim$6.8,  $\sim$7.1 and $\sim$8.0 keV) the
EWs were found to be variable at greater than 99$\%$ confidence. Visual examination of figure 5 suggests the individual absorption line EWs vary in a
non-random manner. In particular, the $\sim$6.8 and $\sim$8.0 keV absorption depths both appear to increase over the first 4 orbits, and follow a visually
similar pattern over the remaining orbits. Since the strong $\sim$7.1 keV absorption line does not follow the same pattern, the implication is of a change
in ionisation state, rather than column density, in the corresponding  velocity component of the (post shock) outflow. 

To test those impressions we searched for correlations between the absorption line EWs and continuum luminosities, using the standard Pearson (linear)
correlation coefficient. The four absorption lines were compared to each other and to the 7-10 keV and 0.3-3 keV luminosities. The only strongly
significant correlation (at $\ga$99$\%$ confidence) was found between the EWs of the $\sim$6.8 and $\sim$8.0 keV absorption lines. Over the whole data set
the  $\sim$6.8 and $\sim$7.1 keV lines were not significantly correlated, while none of the lines showed a significant correlation to the continuum flux.
However, visual inspection of figure 5 does suggest a link of EWs with extreme continum values on an orbit-to-orbit timescale, and similar trends over
several adjacent orbits.  While these findings are not simply understood, the lack of a strong correlation with continua levels overall may indicate a
delayed response to the ionising flux for both the high and lower velocity (putative pre- and post-shock) flow components, or - as indicated below - a
competition between hard ionising radiation and soft Compton cooling flux.

The inter-line correlations are not decisive on the identification of the $\sim$8 keV aborption. While the $\sim$6.8 and $\sim$8.0 keV
absorption depth correlation is expected if both lines are from FeXXV, the similar EWs would require the 1s-2p resonance absorption to be strongly
saturated, inconsistent with its marked variability. In addition, a positive correlation with the most likely alternative candidate for line 4, namely a
high velocity component in Fe Lyman-$\alpha$,  could also arise if the mean ionisation states of the post-shock and pre-shock flow components lie,
respectively, between  Fe XXV/ Fe XXVI and  Fe XXVI/XXVII. Comparison with a photoionisation grid shows that situation implies a difference in
ionisation parameter for the two flow components in the range $\sim$6--8 (Kallman \et\ 2004),   qualitatively consistent with the factor 4 increase in
density of the flow across a strong shock, followed by a further increase as the post-shock flow slows to $\sim$5500 km s$^{-1}$ (Paper I). The continuum 
driving model for high velocity AGN winds (King and Pounds 2003) would predict such an extreme launch ionisation parameter.

\begin{figure*}                                                          
\centering                                                              
\includegraphics[width=12cm, angle=270]{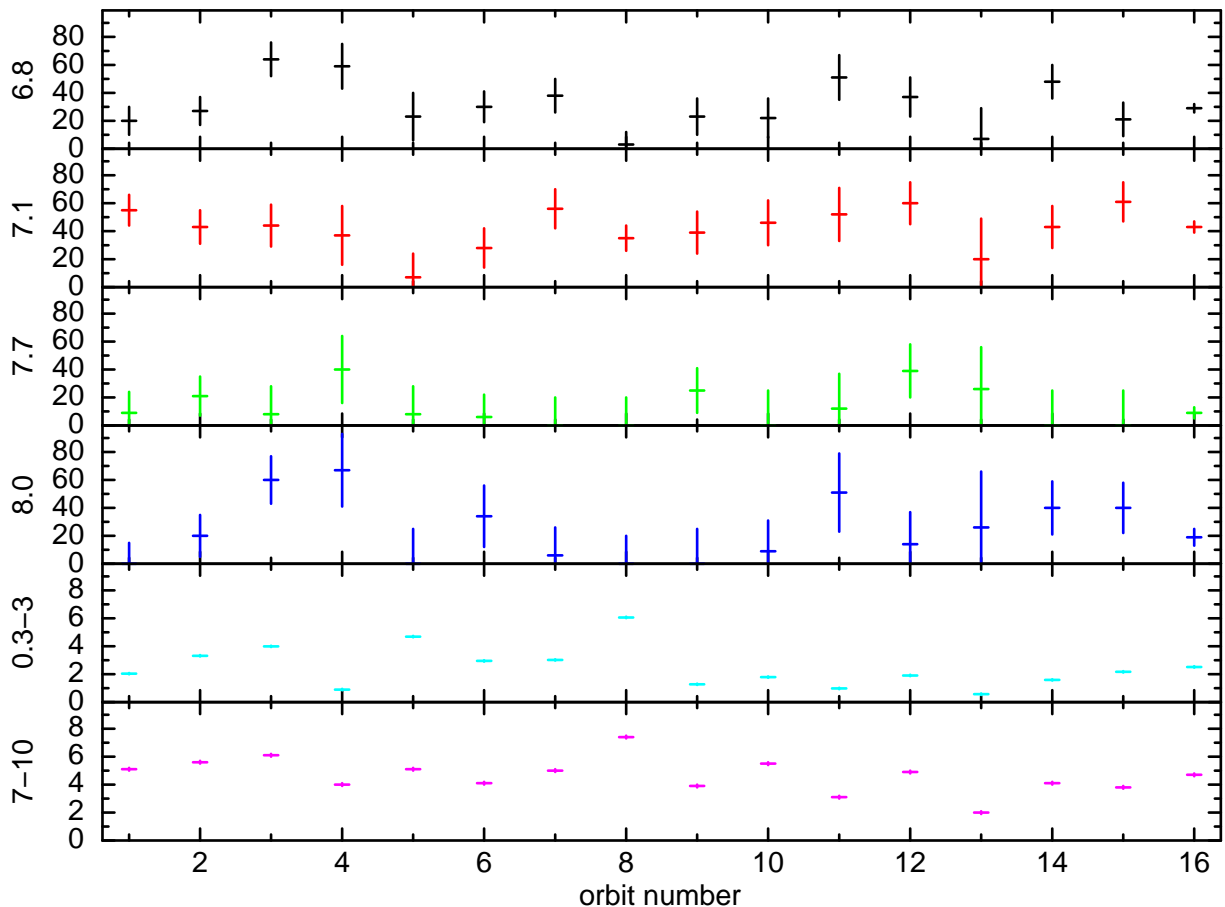}                                                                                               
\caption                                                                 
{Short term variation in the equivalent width of the 4 strongest absorption lines compared with the 0.3-3 keV luminosity (in units of 10$^{41}$ erg
s$^{-1}$) and 7-10 keV luminosity (units of 10$^{40}$ erg s$^{-1}$), obtained by fitting individual orbit spectra to the template derived from modelling
the summed 2009 data}      
\end{figure*} 

\subsection {Individual orbit spectra in 2009}

We review below some individual-orbit spectra with the aim of clarifying the apparent variations in ionisation state of the absorbing outflow, and find
additional orbital timescale changes in emission. Figure 6 shows the best-fit Fe K profiles for the first 8 orbits (rev 1721 to rev 1730) of the 2009 \xmm\ campaign. Orbit-to-orbit differences are  seen
in both absorption and emission structure.

\begin{figure*}
\begin{center}
%
%
\hbox{
 \hspace{1.0 cm}
   \includegraphics[width=5.3cm, angle=270]{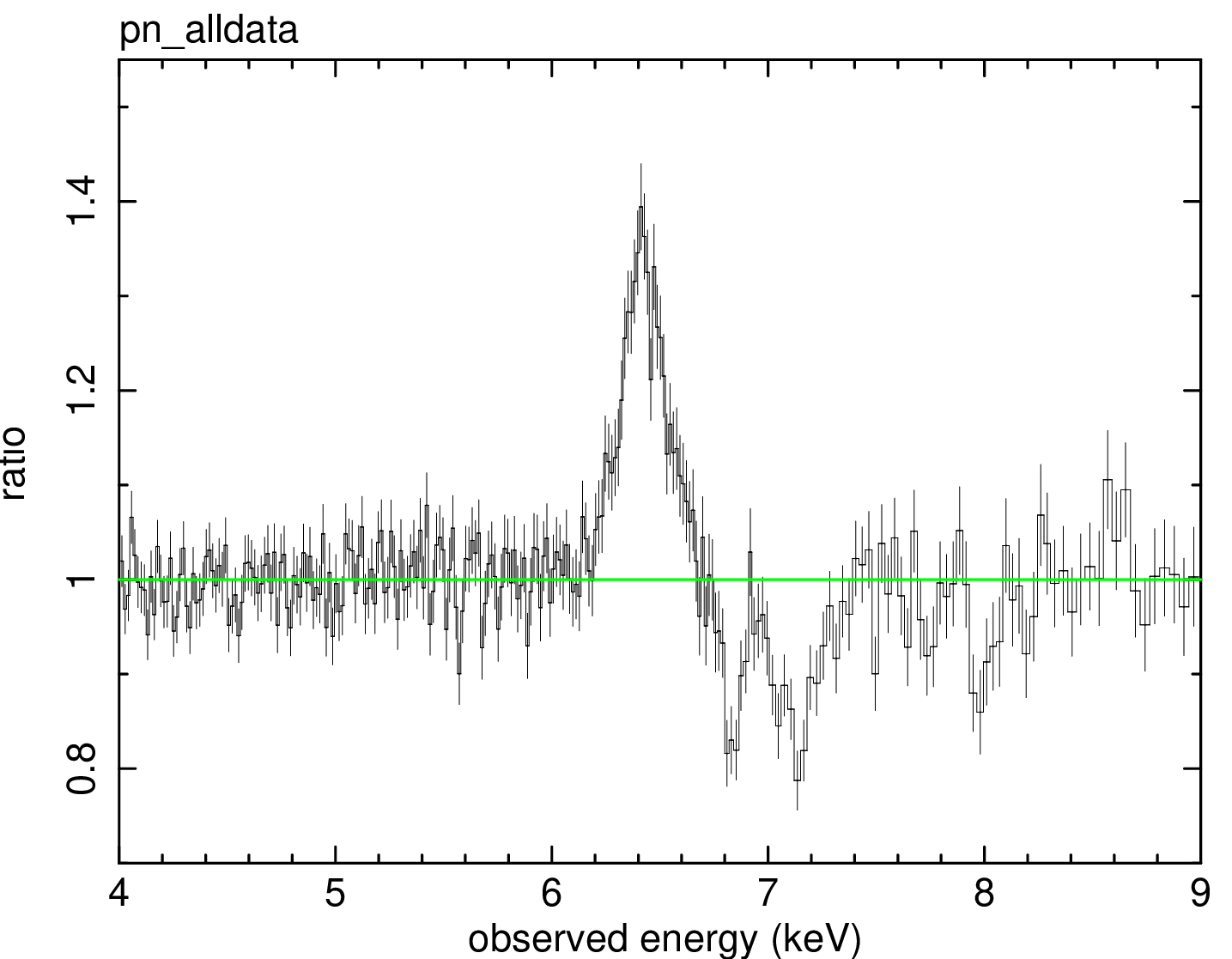}
 \hspace{0.5 cm}
   \includegraphics[width=5.3cm, angle=270]{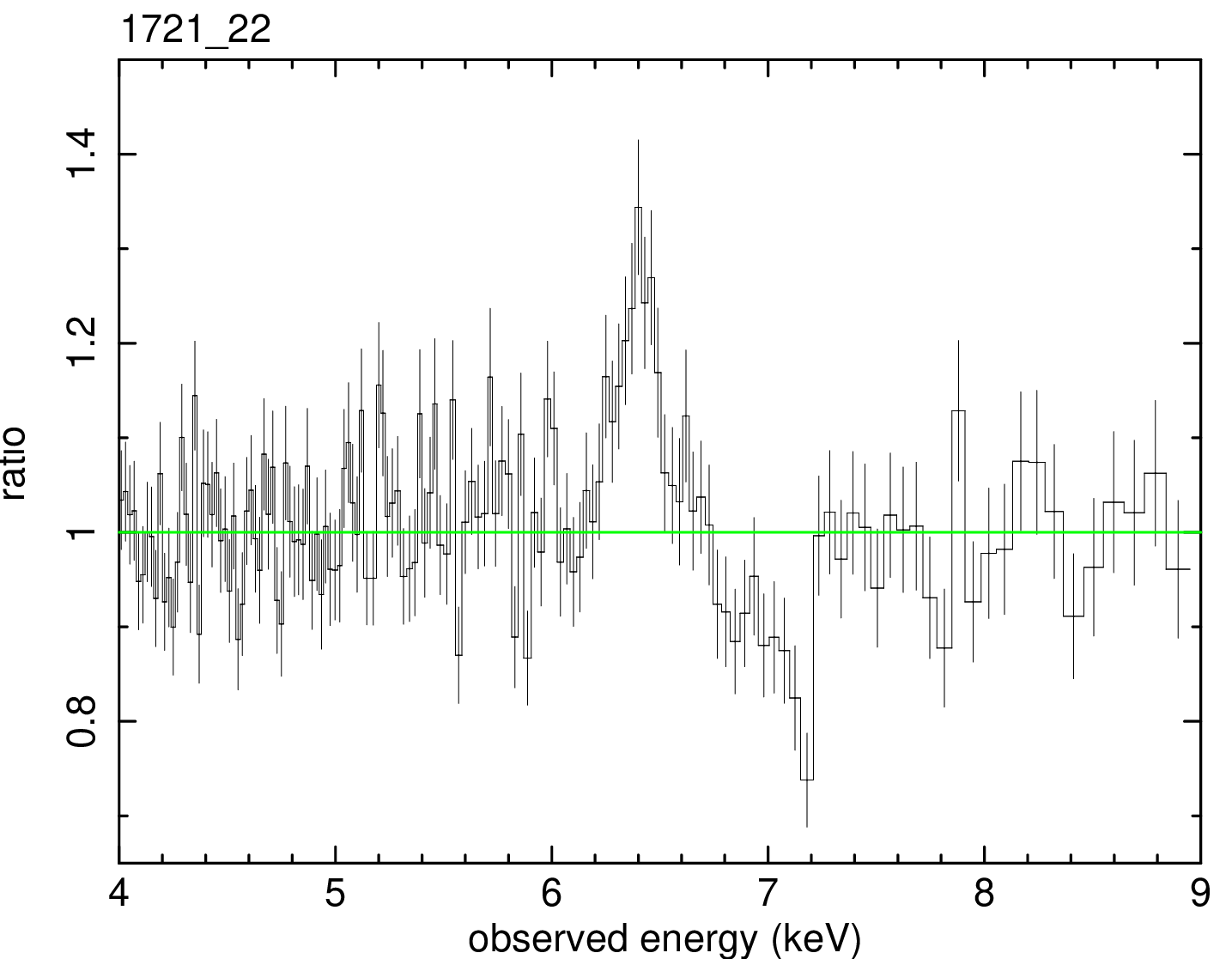}} 
   \hbox{
 \hspace{0.875 cm}
   \includegraphics[width=5.3cm, angle=270]{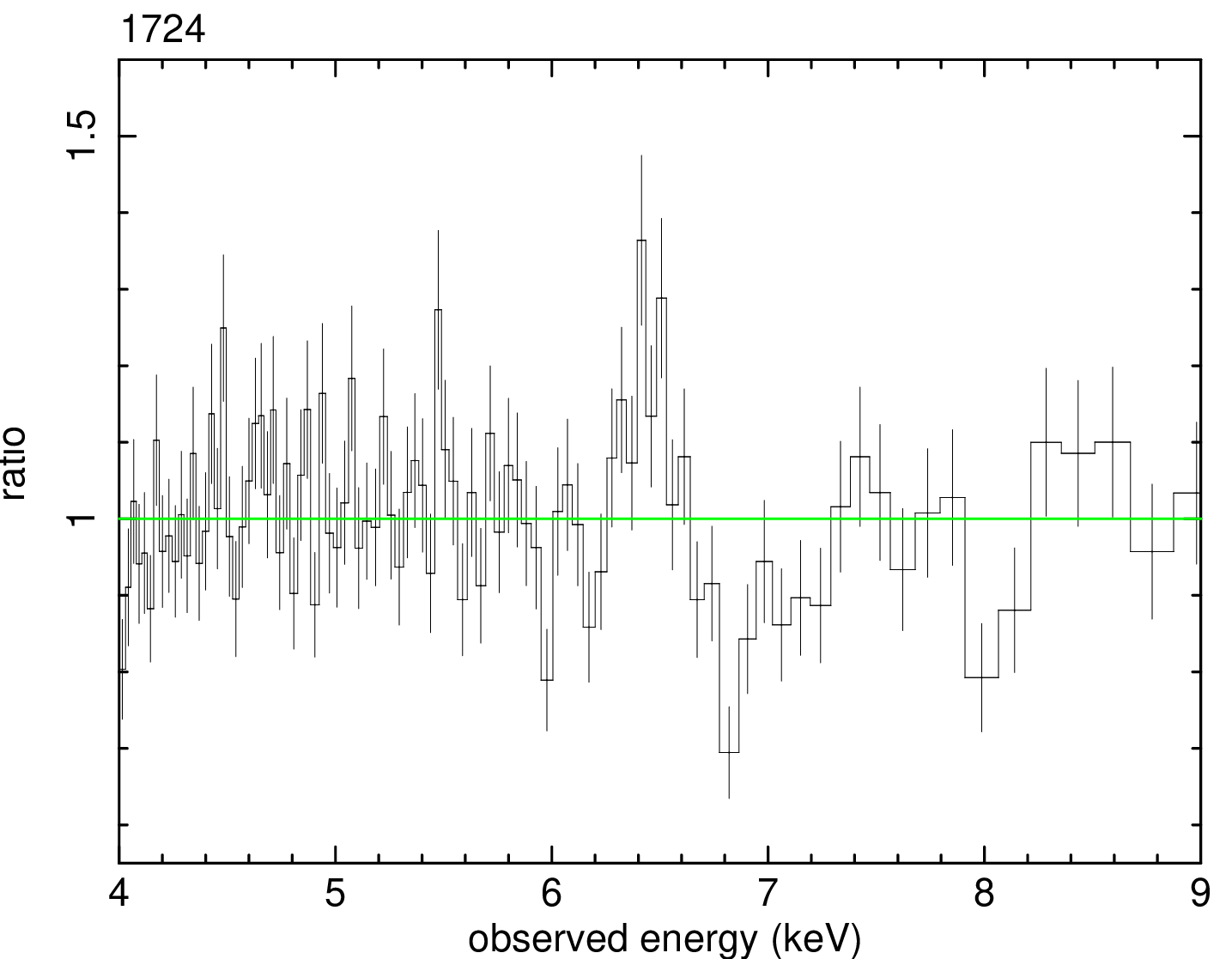}
 \hspace{0.5 cm}
   \includegraphics[width=5.3cm, angle=270]{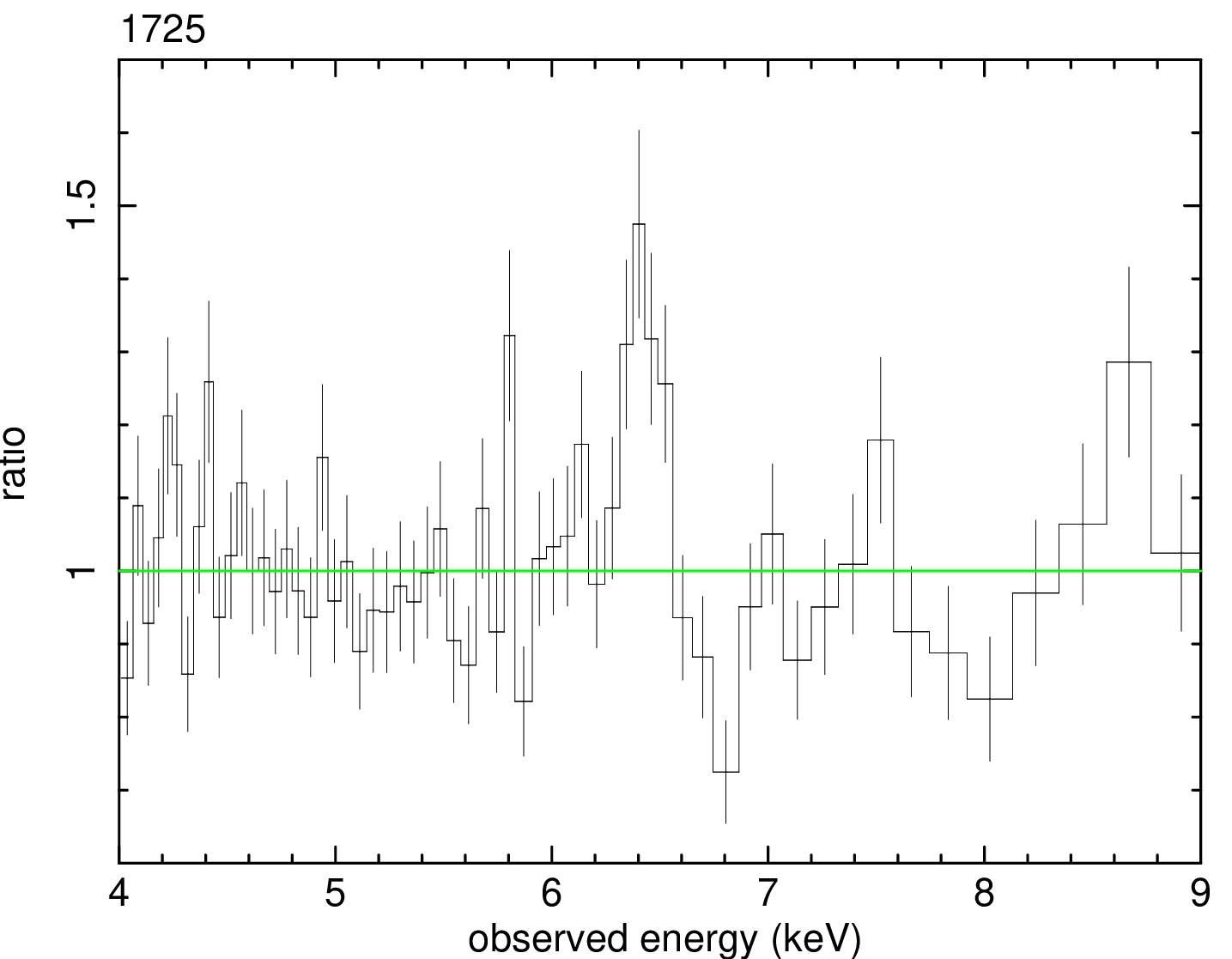} } 
   \hbox{
 \hspace{1.0 cm}
   \includegraphics[width=5.3cm, angle=270]{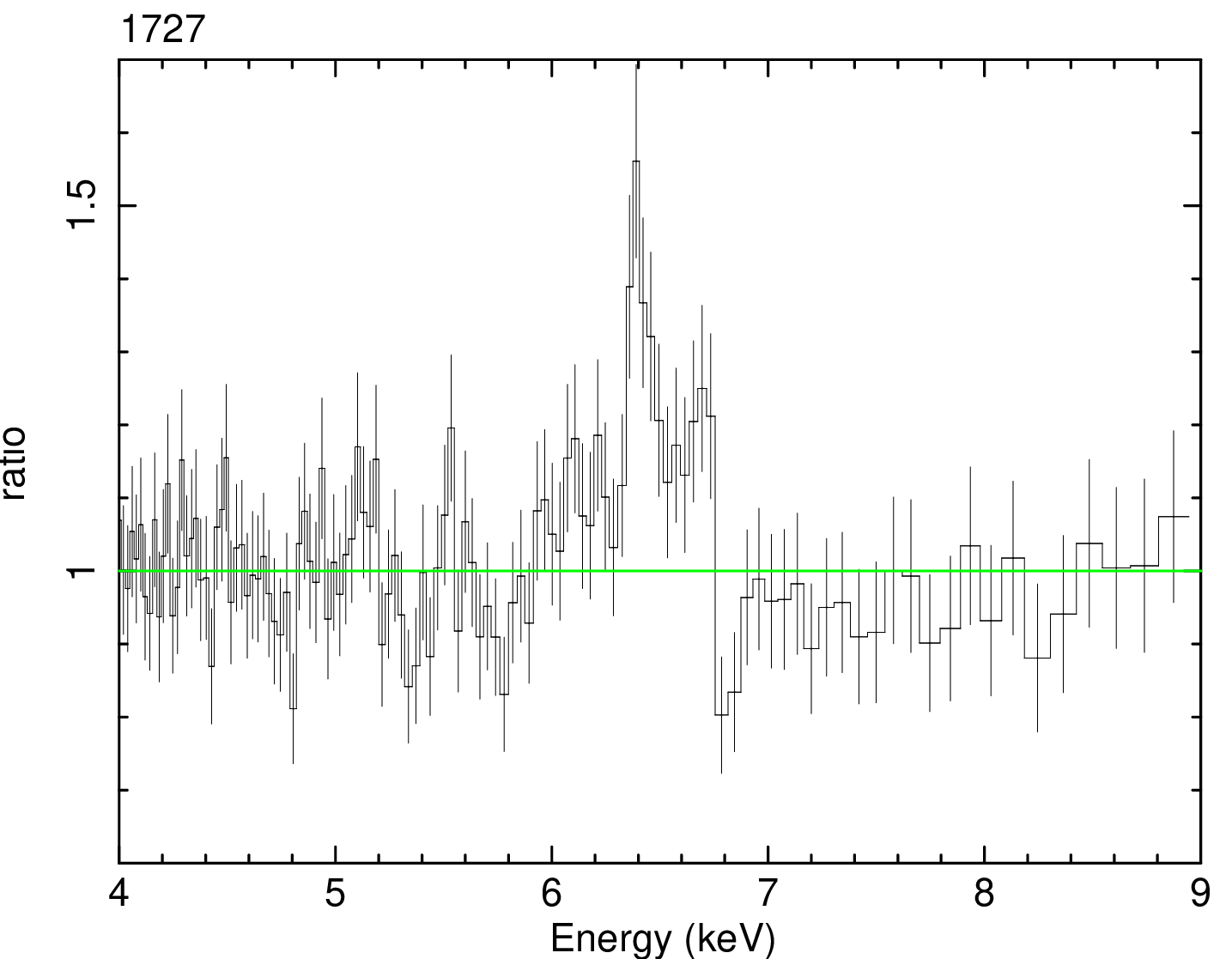}
 \hspace{0.5 cm}
   \includegraphics[width=5.3cm, angle=270]{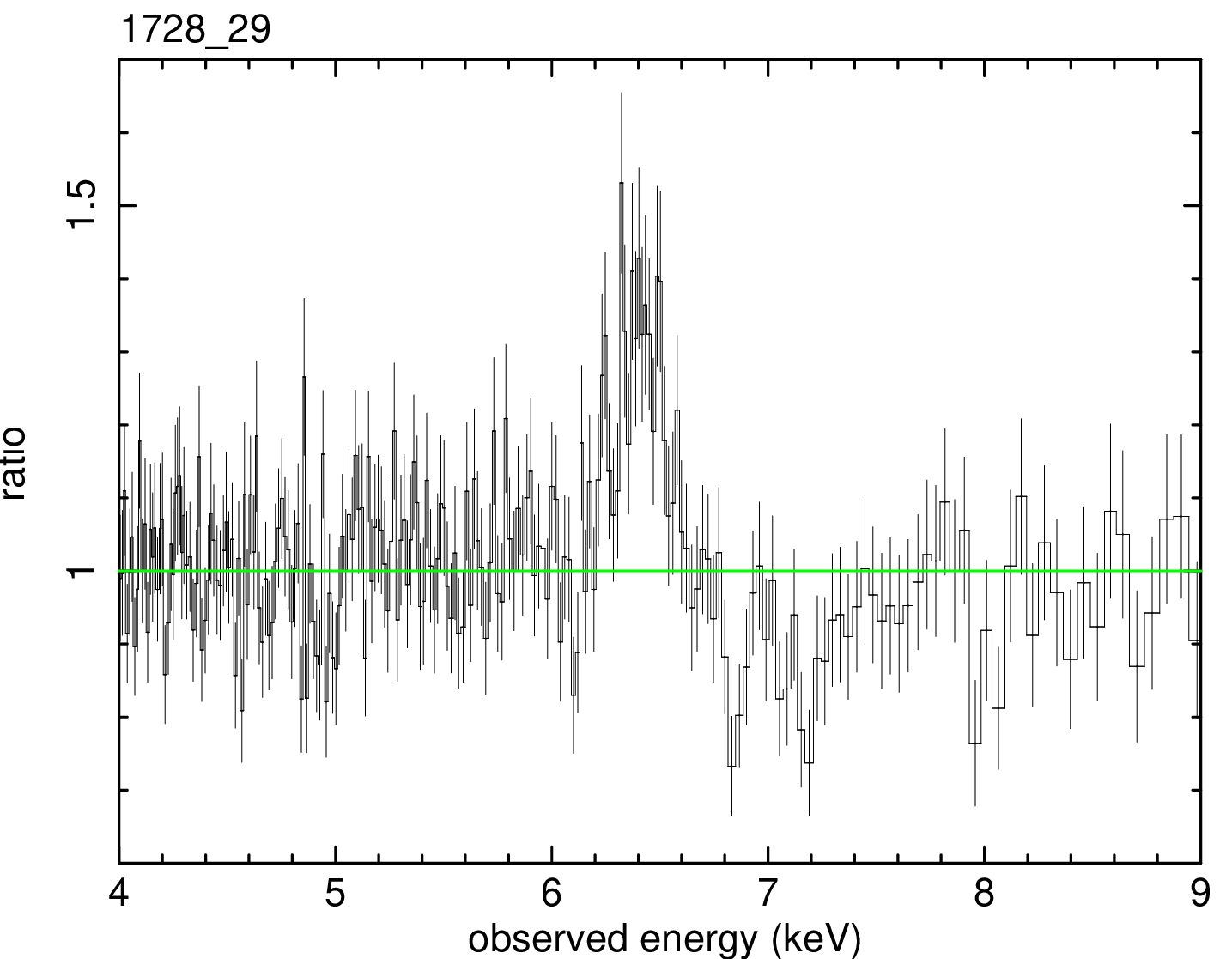} } 
   \hbox{
 \hspace{1.0 cm}
   \includegraphics[width=5.3cm, angle=270]{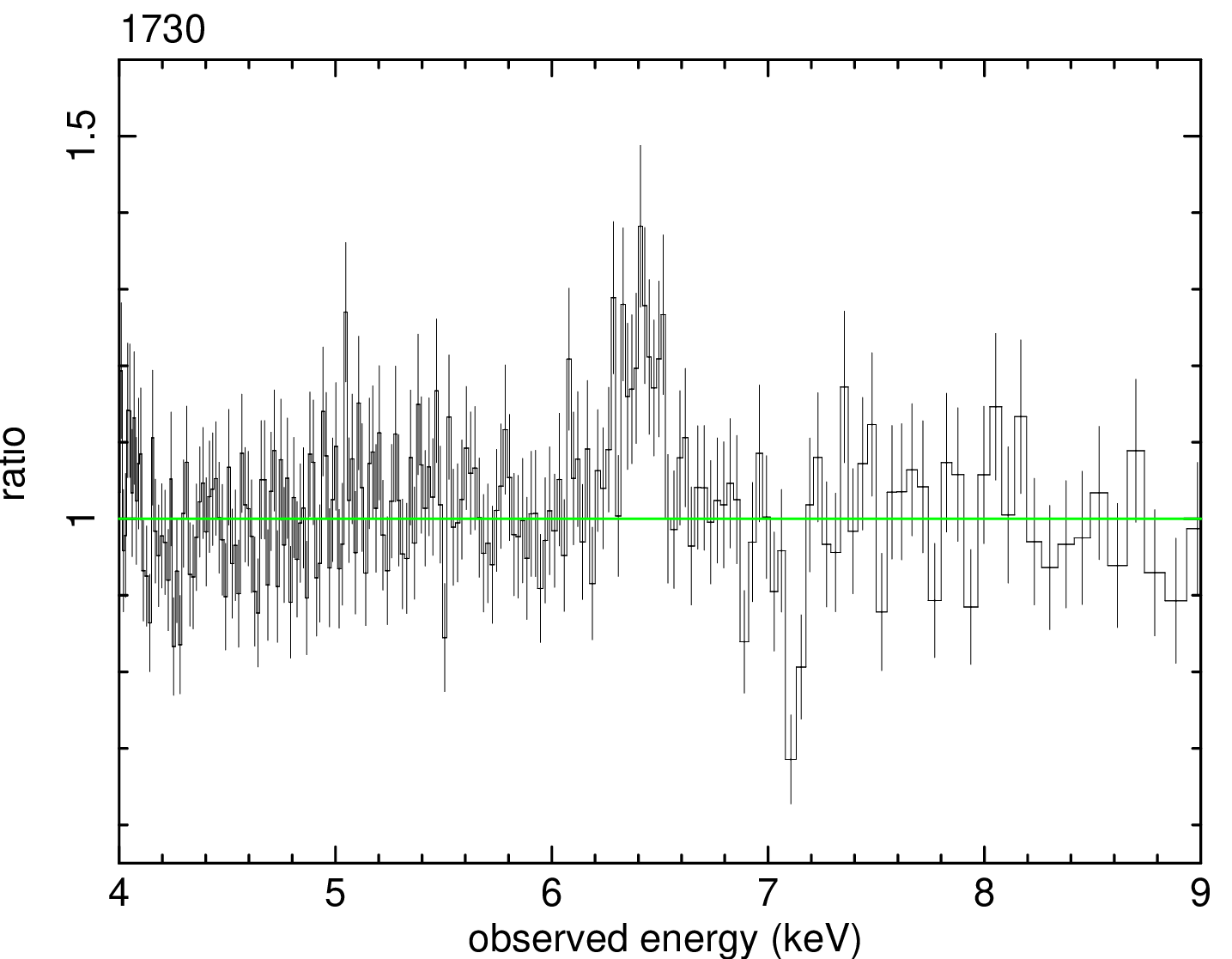}
 \hspace{0.5 cm}
   \includegraphics[width=5.3cm, angle=270]{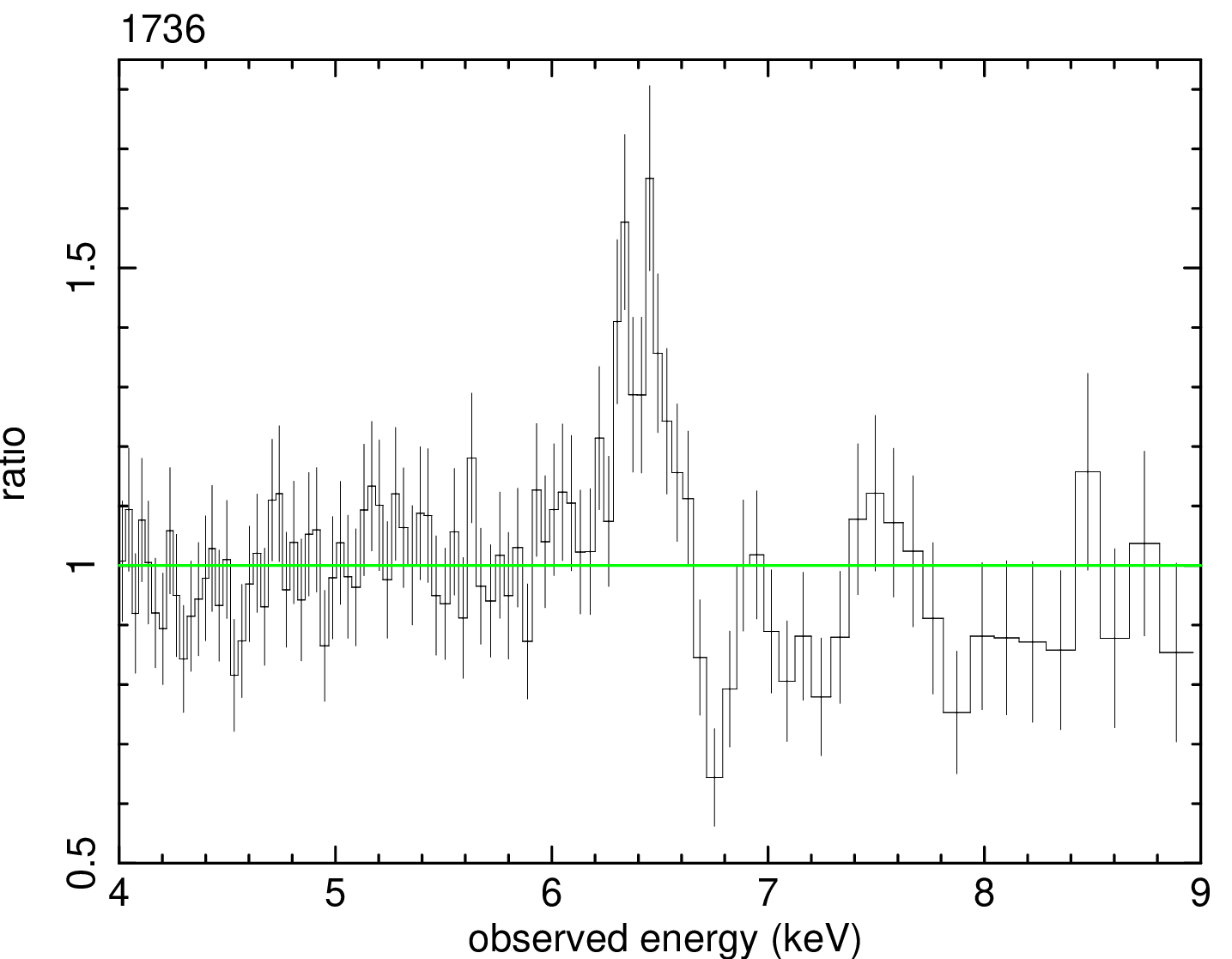} } 
   \end{center} %
\caption{Fe K profiles for the first half of the 2009 \xmm\ campaign (rev 1721 to rev 1730) and a further low flux observation (rev 1736)} %
\end{figure*}

At top left the mean profile for the summed 2009 pn data is reproduced, noting - in particular - similarly strong absorption lines at $\sim$6.8 and
$\sim$7.1 Kev. In contrast, the sum of the first two orbits, revs 1721 and 1722 (similar individual orbit profiles have been added in figure 6 for
improved definition) shows the 7.1 keV line to be dominant. However, by the next observation, rev 1724, the situation has reversed with the 6.8 keV line
now the stronger. Identifying this line pair, as before, with resonance absorption in Fe XXV and Fe XXVI, respectively, indicates the ionisation state of
the relevant outflow component, at v$\sim$5000-7000 km s$^{-1}$, has fallen over that 4 day interval. Figure 5 shows the 7-10 keV luminosity
{\it rising} over those early orbits, which rules out a simple link with  the ionising continuum $\ga$7 keV, unless there is a substantial recombination
time delay in the relevant outflow component after some (unseen) high flux level prior to  the first \xmm\  observation. 

However, figure 5 does show those first 3 orbits to be on a stronger rising trend in soft X-ray flux, the 0.3-3 keV flux doubling from rev 1721 to rev
1724, offering an alternative explanation of the change in ionisation state which is qualitatively consistent with dominant Compton cooling of the
post-shock gas, as proposed in Paper I. The broad range of ionisation parameter over which the He-like ion stage is dominant might then
explain the similar absorption profile being maintained as the overall continuum flux falls in rev 1725. 
The Compton cooling interpretation is supported  by the following (rev 1727) profile, where the 6.8 KeV absorption line remains dominant, and where figure 5 shows the
soft X-ray luminosity to be still higher. The rev 1727 profile also shows a strong emission shoulder to the high energy side of the narrow
Fe K line. We return to considering that emission feature in Section 5.2.

The 6th panel in figure 6, for revs 1728 and 1729 (again similar individual orbit profiles have been added) shows
the ionisation state to have recovered to that of the summed data, while the 0.3-10 keV luminosity has fallen by over a third  from the high  rev 1727
value, with the ionising continuum (7-10 keV) luminosity remaining little changed.
The new trend of increasing ionisation parameter is extended in rev 1730, with the 7.1 keV line now dominant. Although the soft X-ray luminosity has
again increased in rev 1730, figure 5 shows this is the only orbit for which the 7-10 keV luminosity is significantly above
the mean value. We suggest later that the factor $\sim$2 increase in ionising flux in rev 1730, the highest of the campaign, is the probable cause of this 
highly ionised state.

The final panel in figure 6 shows the Fe K profile for rev 1736, where the X-ray continuum has again fallen to a low value, very similar to that of rev 1725. The
absorption spectrum is also very similar to rev 1725, with a dominant line at $\sim$6.8 keV indicating the ionisation state of the $\sim$6500 km s$^{-1}$ flow
component is again  below the 2009 mean.

\subsection{The unusual Fe K profile of rev 1727}

The Fe K profile for rev 1727, coincident with an
unusually  strong soft X-ray flux, is particularly striking, and notably different from that for the summed 2009 data. While only the $\sim$6.8 keV is absorption line 
can be seen, P Cygni emission is clearly visible on the low energy shoulder of that line. (\footnote {An alternative possibility not explored further here is that the
diskline parameters have changed to give a more sawtooth profile.})  
Adding positive and  negative Gaussians to model the P Cygni structure (figure 7) yields a significantly improved fit, with $\Delta$$\chi^{2}$ of 21 for 5 fewer
d.o.f.  The absorption line at 6.81$\pm$0.02 keV is unresolved (and then fixed at the pn resolution) while the emission component has a 1$\sigma$ width of 115$\pm$38 eV and energy
6.71$\pm$0.05 keV. The two components are of similar EW, indicating a substantial covering factor, with an inflection point at 6.76$\pm$0.04 keV,
consistent with the rest energy of the Fe XXV 1s-2p resonance transition  in the rest frame of \ngc. Although not well constrained, the emission line
width is also compatible with a wide angle outflow for the $\sim$6000 km s$^{-1}$ velocity component, associated in our model with the post-shock gas.
We note the potential of higher quality data of such features in mapping the geometry of the corresondingly flow components. 

To the low energy side of the Fe line in the rev 1727 data both positive and negative spectral features can be seen, each of which is
marginally significant, with $\Delta$$\chi^{2}$ of 6 to 8. Those at $\sim$5.1 and $\sim$5.5 keV might be related to 
to the  emission line reported by Turner \et\ (2010); see also Section 6.  Similarly, absorption lines indicated at $\sim$5.35 and $\sim$5.75 keV
could correspond to the anticipated in-fall of cool, shocked gas with  velocities of order 0.2c and 0.11c, respectively. However, unlike the
absorption line at $\sim$6.8 keV, these lower energy features in the rev 1727 profile do not appear in the spectrum summed  over the whole 2009 observation, suggesting they are short lived
features or outlying statistical fluctuations.

\begin{figure}                                                           
\centering                                                              
\includegraphics[width=6.28cm, angle=270]{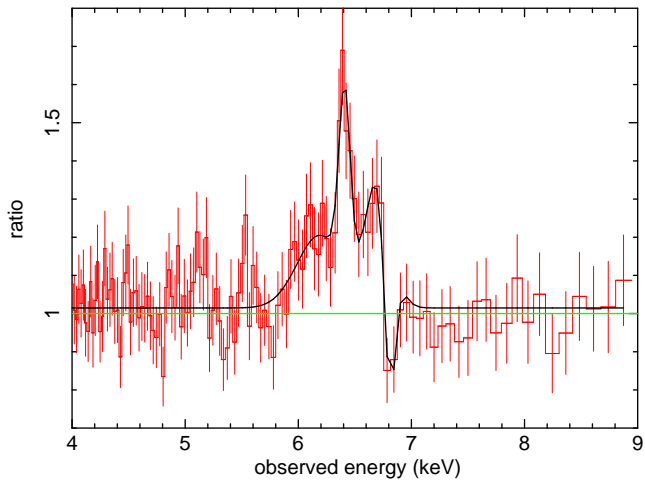}                                                                                          
\caption                                                                 
{The Fe K profile for rev 1727, coinciding with a strong soft X-ray (and UV) flux, is notably different from that for the summed 2009 data, with only the lower
energy absorption line at $\sim$6.8 keV clearly seen and visual evidence for P Cygni emission. The figure includes positive and negative
Gaussians to model that profile}      
\end{figure}

\section{Another look at the 2001 and 2002 \xmm\ pn data}

Pounds \et\ (2004) previously reported on the \xmm\ observations of \ngc\ in 2001 and 2002, when the source was in high- and extended low-flux states,
respectively. The EPIC spectrum from 2001 (rev 0263) showed a single absorption line at $\sim$7.1 keV in both pn and MOS data, with alternative
identifications of Fe XXVI Lyman-$\alpha$ and Fe XXV 1s-2p resonance transitions indicating a corresponding outflow velocity of $\sim$6500 or $\sim$16500
km s$^{-1}$. A comparison with the 2009 summed data template, with only the direct power law and absorption line depths free (a prior check showing the narrow Fe K
emission line flux, hence cold reflection, to be only marginally higher),  yielded an excellent 3-10 keV fit ($\chi^{2}$ = 582/610). Figure 8 (top panel)
reproduces the ratio plot of the pn data, confirming  a dominant, broad absorption line at $\sim$7.1 keV.

\begin{figure}                                                          
\centering                                                              
\includegraphics[width=6.28cm, angle=270]{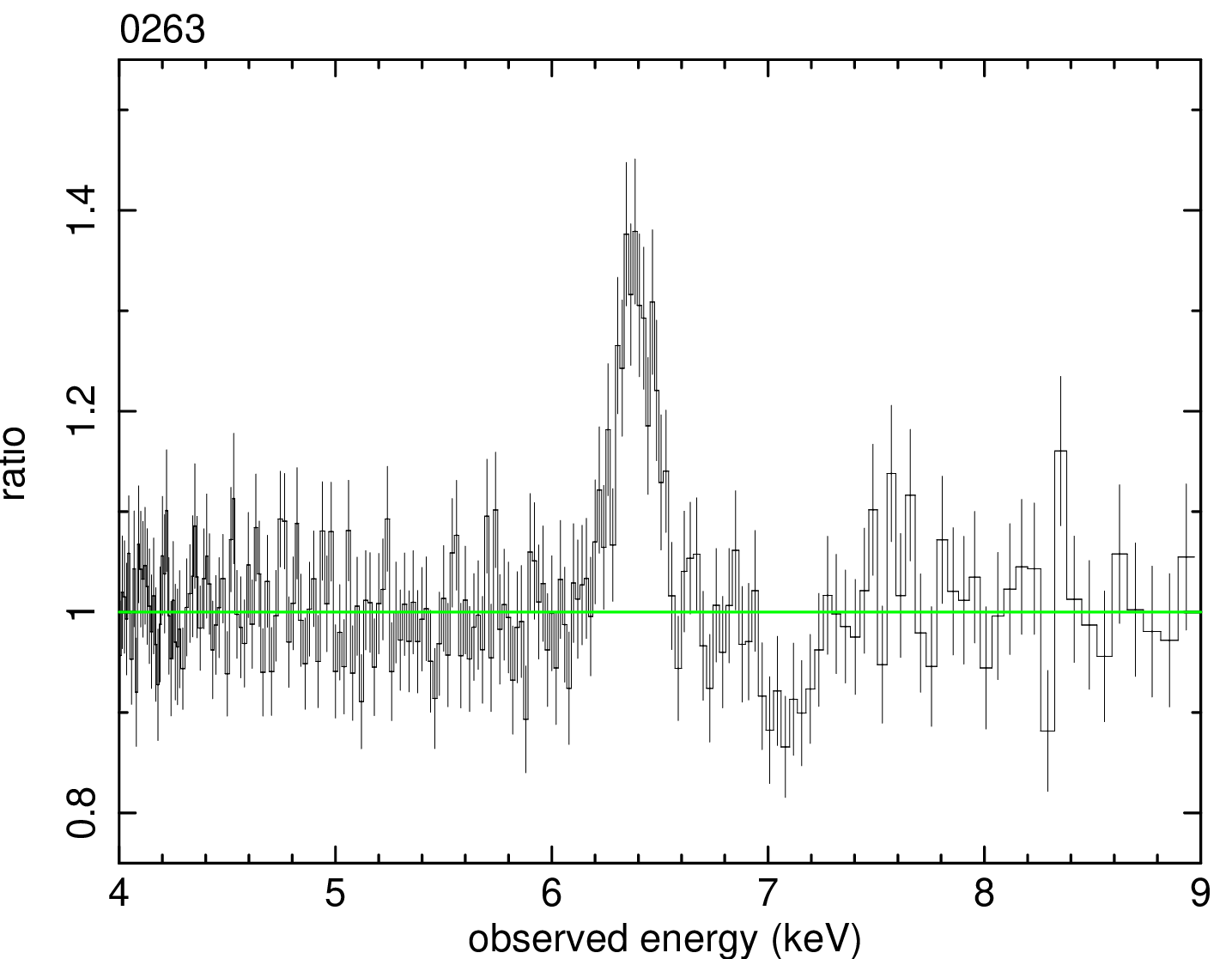}          
\centering                                                              
\includegraphics[width=6.28cm, angle=270]{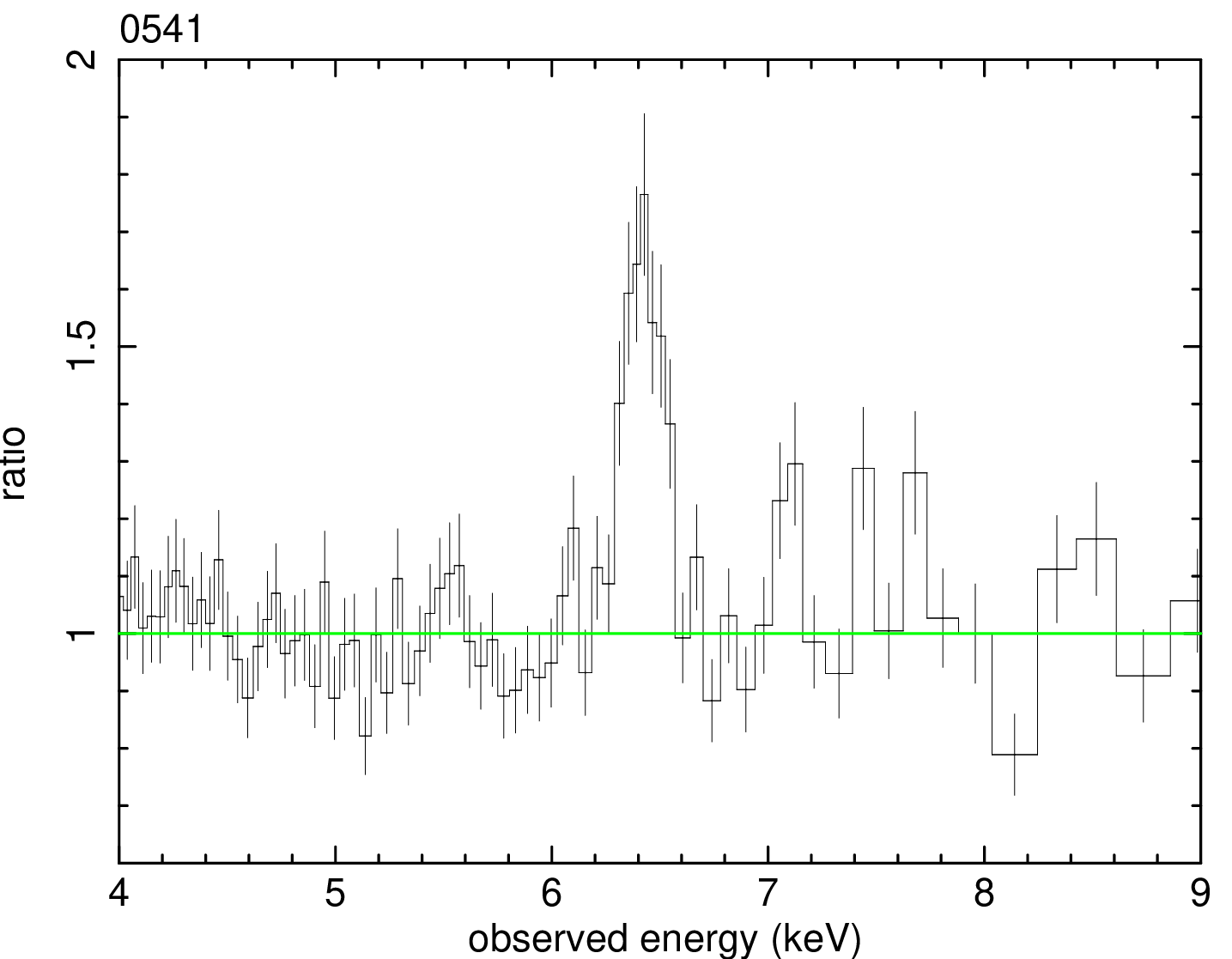}                                                                                      
\caption                                                                 
{The Fe K profile for the 2001 \xmm\ observation shows a single broad, blue-shifted absorption line with excess flux to the low energy side of the 
absorption line hinting at a P Cygni line arising from an outflow seen in Fe XXVI Lyman-$\alpha$. The Fe K profile for rev 0541 is dominated by a narrow
emission line at $\sim$6.4 keV,  with previously reported features in emission at $\sim$5.54 keV and absorption at $\sim$8.10 keV also found to be
significant}      
\end{figure}

A fit to the 2002 \xmm\ observation (rev 0541), taken after a 2-week period of unusually low X-ray emission (Uttley \et\ 2003), was seen in Section 3  to
be very similar to that of the lowest flux orbit (rev 1739) in 2009. Figure 8 (lower panel) shows the Fe K profile from rev 0541, derived at above, dominated by a narrow Fe
K-$\alpha$ emission line at $\sim$6.43$\pm$0.01 keV. Once again, the line is resolved, with  width  $\sigma$=85$\pm$10 eV. A statistically significant
detection of the Fe K-$\beta$ line at $\sim$7.1 keV finds an EW a factor $\sim$5 lower than that of Fe K-$\alpha$.

The rev 0541 fit is further improved with the addition of a narrow emission line at 5.54$\pm$0.02 keV ($\Delta$$\chi^{2}$ of 10/2) and an unresolved absorption line at
8.10$\pm$0.1 keV ($\Delta$$\chi^{2}$ of 8/2).  The former feature lies close to that reported at 5.44$\pm$0.03 keV in \suzaku\ data  and attributed to
spallation of Fe into Cr (Turner and Miller 2010).  The high energy absorption line was picked out in the survey for high velocity flows in AGN spectra
from the \xmm\ archive search by Tombesi \et\ (2010), and identified there with Fe Lyman-$\alpha$ for a blue-shifted velocity of $\sim$0.13$\pm$0.01c.

In the context of alternative identifications of the $\sim$8 keV absorption line in the present paper, it is interesting to note that no absorption line
at $\sim$6.8 keV is apparent in the rev 0541 spectrum, while being consistent with the  putative high speed wind being very
highly ionised, becoming more visible in the Fe XXVI Lyman-$\alpha$ line at low ionising fluxes.

In summary, a re-examination of the 2001 data strengthens the earlier interpretation of the line observed at $\sim$7.1 keV as arising from
Fe XXVI resonance absorption from matter outflowing at $\sim$6500 km s$^{-1}$. In the context of our analysis of the more comprehensive absorption data
from the 2009 \xmm\ spectra, that finding suggests the post shock flow is persistent, with an ionisation state changing from dominantly Fe XXV to
dominantly Fe XXVI as the ionising flux increases to the levels seen in rev 0263 (and rev 1730).
The similarity of the narrow FeK emission line flux and (resolved) line width in the 2001, 2002 and 2009 \xmm\ observations suggests 
the longer-term persistence of the weakly ionised matter responsible for the Fe K line fluorescence and associated continuum reflection. If that matter is
linked to the post shock flow in \ngc, as suggested here, the important implication is that the high speed wind is also a persistent feature.

\section{Discussion}

The 600 ks observation of \ngc\ in 2009 has provided some of the richest X-ray spectra to date for studying the outflow properties in an AGN. In addition
to the increased sensitivity afforded by the unusually long  observation, comparison with previous observations from \xmm, \chandra\ and \suzaku\ (Collinge \et\
2001, Pounds \et\ 2004, Steenbrugge \et\ 2009, Lobban \et 2011)  suggests that our 2009 campaign benefited from an overall timescale compatible with intrinsic variability in the flow parameters.

In Paper I we identified 3 broad velocity regimes in the outflow in line-of-sight to \ngc, referred to there as the low ($\leq$1000 km s$^{-1}$), 
intermediate ($\sim$3000-9000 km s$^{-1}$) and high velocity ($\sim$0.1c) flows, with the low and intermediate velocities well populated by absorption
lines from a wide range of metal ions from C to Fe. The highest velocity component was only detected - at relatively low confidence - in Fe K
absorption.   

Paper I then explored the possibility that the three velocity regimes represent different stages in a shocked high speed wind (King 2010), where the intermediate
velocity/intermediate ionisation outflow corresponds to the immediate post-shock gas and the low velocity/low ionisation absorption to matter building up
ahead of the contact discontinuity. In a second paper (Paper II), we proposed that a low velocity absorption component observed in RGS spectra near the
line cores of several broad resonance emission line components had a separate origin as self-absorption in the limb-brightened shell of shocked gas. 

In the King (2010) model a high velocity ionised wind collides with the ISM of the host galaxy, resulting in a strong shock. The gas density increases by
a factor $\sim$4 at the shock front, and the velocity drops by the same factor. Beyond this (reverse, adiabatic) shock, the flow is further compressed in
a relatively thin, cooling region, while the velocity slows to low values at the interface with the ISM. Strong Compton cooling by  the AGN soft X-ray/UV
radiation leads to a fairly rapid transition  between the  immediate post-shock regime and the much slower and compressed state near the contact
discontinuity.

The primary aim of the present paper has been to examine the complex Fe K emission and absorption profile and attempt to resolve and identify spectral
features carrying information on the highly ionised outflow and re-processing of the primary X-ray continuum. We have employed two methods of analysis, 
first with Gaussian fitting of the ratio for
the summed pn data profile to a simple power law baseline, and then by a more physical modelling in {\it Xspec} where the continuum is fitted with a
weakly ionised reflection component (consistent in strength and ionisation state with the observed narrow Fe K fluorescence line), and a partially 
covered direct power law. 

We find the absorption line structure from the two methods to be in good agreement. Detection of 3 (or 4) blue-shifted absorption lines provides evidence 
for two velocity components in the highly ionised outflow, with a factor $\sim$4
difference  in velocity as predicted for a shocked flow, although the high velocity component is probably blended with the Fe XXV 1s-3p line. 
An apparent broadening of the absorption line at $\sim$7.15 keV
(identified with Fe Lyman-$\alpha$ and a outflow velocity range of $\sim$5000-9000 km s$^{-1}$) is found in both analysis methods, notwithstanding the
proximity of the Fe K absorption edge in the broad band fit. The similarity to the broad absorption trough seen in the RGS data for oxygen Lyman-$\alpha$
(Paper II) again indicates matter of differing densities in a co-moving, decelerating flow.  

Our analysis fits the excess flux in the $\sim$5.5-7 keV interval with a narrow Fe K emission line and a weak relativistic red wing. We note that the higher
resolution (but lower sensitivity) \chandra\ HETG spectrum in Lobban \et \ (2011) finds a `narrow' line component almost a factor two broader than in our
analysis, although a second fit with an unresolved core underlines the limitation of such simple Gaussian fits. What is interesting is that both Lobban
\et\ and our present analysis find a substantial Fe K flux arising from weakly ionised matter with a projected velocity dispersion of order $\sim$$10^{4}$
km s$^{-1}$ (FWHM). An appealing interpretation is that this component arises from continuum X-ray flux being scattered from the post-shock outflow,
the velocity width implying a wide angle flow.

Comparison of individual orbit spectra with that of the summed data finds the ionisation state of the post shock flow, as determined by the ratio of line opacities of  Fe XXVI to Fe XXV, is seen to fall over the first 5 orbits of the 2009 
campaign, while the 7-10 keV luminosity, a proxy for the ionising flux affecting the Fe
K shell, varies very little. On the other hand, the 0.3-3 keV luminosity increases from rev 1721 to rev 1727 by a factor $\sim$2.3, and simultaneous UV
flux measurements show a similar increasing trend (Alston \et\ 2012), suggesting the observed ionisation change is responding to enhanced cooling, with a 
lower gas
temperature and corresponding increase in recombination rate.  Following that reasoning we can then understand the subsequent reversal to a higher
ionisation parameter for the same flow component in revs 1728 and 1729, where the 0.3-3 keV luminosity has fallen back by $\sim$ a third.
If confirmed, the above interpretation of the absorption line variations over the first half of the 2009 \xmm\ campaign provides independent support for
the shocked outflow model discussed in Paper I, where strong cooling of the post-shock flow is a key to the ionisation-velocity correlation seen in the RGS
spectra.

However, while the soft X-ray flux has again increased by rev 1730, that observation sees the largest ionising flux of the campaign, with the 7-10 keV
luminosity a factor $\sim$2 higher than in rev 1729. The dominant 7.1 keV absorption line, also seen in the similarly high flux observation of 2001 
(rev 0263), indicates a still higher ionisation parameter for the $\sim$5000-7000 km s$^{-1}$ flow component, with photoionisation now dominating over
Compton cooling.  If the line ratio change from rev 1729 to rev 1730 is indeed due to an
ionising pulse, the $\sim$2 day timescale is much shorter than for an ionising front travelling at the local sound speed ($\leq$1000 km
s$^{-1}$)  through the shell of post-shock gas, estimated in Paper I to be of thickness $\sim$$10^{16}$ cm. Alternatively, for an R-type ionisation front,
moving into a low density medium, the response time will be  governed by the recombination time. For an assumed post-shock temperature of $\sim$0.1 keV
(Paper I), a recombination  coefficient of $3.5\times10^{-11}$  (Verner and Ferland 1996) would then relate the measured recombination timescale of
$\sim$$2\times10^{5}$ s to a mean (electron)  density of $\sim$$1.5\times10^{5}$ cm$^{-3}$.

That value is consistent with the measured absorption column of N$_{H}$$\sim$$2\times10^{21}$ cm$^{-2}$ from the ionised Fe absorption line EWs, and a shell
width (Paper I) of $\sim$$10^{16}$ cm.
Comparison with a density of  $\sim$$5\times10^{5}$ cm$^{-3}$ deduced from variability in the oxygen RRC (Paper II) would support the interpretation of
co-moving ionised O and Fe flow components in pressure equilibrium with a temperature of 0.03 keV (from RRC widths) for the less ionised gas. A further implication of the
inhomogeneous co-moving flow is that the ionisation  parameters will differ in the ratio of the densities. While that ratio is somewhat less than the
factor $\sim$10 predicted for dominant O VIII and Fe XXV ion stages, the agreement is not unreasonable given the approximate nature of the above estimates.

\subsection{A transient high velocity wind in \ngc}

In Paper I we noted that absorption in the immediate post-shock gas, observed at velocities up to 
v$\sim$9000 km s$^{-1}$ and ionisation parameter to log$\xi$$\sim$3, implied the presence of a still higher velocity wind, with v$\sim$36000 km s$^{-1}$ and log$\xi$$\sim$3.6. Although the
evidence presented in Paper I for such a highly ionised wind was marginal, we noted the Tombesi \et\ (2010) report of the detection of an outflow at
$\sim$0.13c (v$\sim$39000 km s$^{-1}$) in the 2002 \xmm\ observation of \ngc. We have repeated the 2002 data analysis in Section 6, finding the $\sim$8
keV absorption to be the only significant absorption line in the Fe K spectral region, with its detection after 2 weeks of extreme low continuum flux
being consistent with an increased fraction of Fe XXVI in a highly ionised wind. 

The \chandra\ and \suzaku\ observations of \ngc, some 6 months prior to the 2009 \xmm\ campaign, yielded absorption
spectra showing no strong evidence in the soft X-ray band for outflow velocities $\ga$1000 km s$^{-1}$  (Lobban \et\ 2011). Although the \chandra\ spectra
are somewhat less sensitive, it seems clear that the soft X-ray opacity at $\sim$3000-9000 km s$^{-1}$ was significantly weaker than 6 months later. On
the other hand, the simultaneous \suzaku\ spectra did show a pair of strong absorption lines at $\sim$6.8 and $\sim$7.1 keV, interpreted - as here - with
He- and H-like Fe, with a projected outflow velocity of $\sim$5000-7000  km s$^{-1}$, but found no evidence for a higher velocity wind. 

An obvious way of reconciling the 2008 and 2009 data is that the high
velocity wind was weaker during the earlier observation, leading to a lower opacity in the post-shock flow. The flow speed and recombination timescales we
have deduced (Paper II, and herein) would readily allow for that different picture over a 6-month interval.   In particular, a denser (more massive)
post-shock  flow evidenced by the strong $\sim$3000-7000 km s$^{-1}$ absorbing flow seen in the RGS spectrum of 2009  would support an assumption that the
fast wind was also blowing more strongly in 2009 than during the \chandra\ and \suzaku\ campaign.

Emission from the pre-shock wind might be expected as a more persistent excess to the red wing of the absorption line at $\sim$8 keV. However, for a 
spherical shell moving
at v$\sim$0.13c, any such feature would probably be correspondingly broad and difficult to detect. A near-sided conical flow as indicated in the RGS data (Paper II)
would give a less broad feature and it is interesting to note the excess near $\sim$7.8 keV in figure 4. Adding a positive Gaussian to that fit, with a
width $\sigma$$\sim$0.2 keV, matches  that excess  and also enhances the absorption line depths at $\sim$7.75 and $\sim$8 keV. The corresponding forward
cone semi-angle is $\sim$45$\deg$, similar to that deduced from the RGS spectral analysis in Paper II and optical imaging of \ngc\ (Christopoulou \et\
1997).

In general, we might expect the putative, highly ionised fast wind in \ngc\ to be intermittent, with the required super-Eddington condition for continuum
driving (King and Pounds 2003) varying on the short (viscous or thermal) timescale of the inner disc. In that context it is important to note  the very high column density
required to detect a blue-shifted Fe K absorption line in a low redshift source such as \ngc, while radial expansion could rapidly render undetectable the
line-of-sight column density in an initially optically thick transient outflow. 
Furthermore, the absence of the $\sim$8 keV absorption line in rev 1730 (and rev 0263) is consistent with the line-of-sight component of the putative pre-shock
wind (seen in FeXXVI Lyman-$\alpha$) being more completely ionised by the stronger continuum $\ga$7 keV.

\section{Summary}

Combining the pn data over the 600ks \xmm\ observation of \ngc\ shows a complex Fe K profile which can be resolved into both narrow and broad (red wing)
emission components and several absorption lines. The narrow Fe K emission line has a mean energy consistent with fluorescence from weakly ionised
matter and a line width consistent with scattering from the outflow seen in absorption.  

Four blue-shifted absorption lines can be identified with Fe XXV and Fe XXVI resonance transitions consistent with line-of-sight outflow velocities
separated by a factor $\sim$4, as expected for a fast highly ionised wind being shocked on impact with the ISM or slower moving ejecta.

Comparing Fe K profiles for individual orbits reveals significant changes on a few-day timescale, with the clear evidence
for variations in the ionisation state of the post shock flow. While no strong correlation of absorption with continuum flux levels is found for all 15 observations
over 45 days, shorter term trends strongly suggest a link of ionisation state with the soft X-ray (and UV) flux and - for the brightest orbit (rev 1730) - with 
a factor 2
increase in hard X-ray flux. While the former link is important in offering support for the existence of shocked gas which is being cooled by disk
photons, the rapid response to a pulse of ionising radiation can be interpreted in terms of a recombination timescale which yields an electron density of 
$\sim$$1.5\times10^{5}$ cm$^{-3}$. Comparison with the somewhat higher density derived in Paper II from the RGS spectral analysis is consistent with
two ionisation components in a co-moving flow.

Finally, we confirm an absorption line at $\sim$8 keV in the 2009  summed spectrum, at a
similar energy and EW to that found in the 2002 data  as part of the Tombesi \et\ (2010) survey, and identified there with an outflow velocity of
$\sim$0.13c. In our more extended 2009 observations we find evidence that the $\sim$8 keV line is variable, being stronger when the flux level is lower.
The implication is that the fast wind is very highly ionised at moderate or high continuum flux levels, with Fe predominently  in Fe XXVII. Recombination
to Fe XXVI (see possible RRC in figure 1) would then increase the wind opacity when the ionising flux level fell. We recall that a  fully ionised wind is
quite likely in the continuum driving model of Black Holes Winds discussed by  King and Pounds (2003).

\section*{ Acknowledgements } 
The results reported here are based on observations obtained with \xmm, an ESA science mission with instruments and contributions directly funded by ESA
Member States and the USA (NASA). We acknowledge illuminating discussions with Andrew King.


\begin{thebibliography}{}
\bibitem{} Alston W.N., Vaughan S., Uttley P., \ 2012 MNRAS, to be submitted
\bibitem{} Arnaud K.A., \ 1996, ASP Conf. Series, 101, 17
\bibitem{} Cappi M., \ 2006, Astron. Nachr., 327, 1012
\bibitem{} Christopoulou P.E., Holloway A.J., Steffen W., Mundell C.G., Thean A.H.C., Goudis C.D., Meaburn J., Pedlar A., \ 1997, MNRAS, 284, 385
\bibitem{} Chartas G., Brandt W.N., Gallagher S.C., Garmire G.P., \ 2002, ApJ, 569, 179
\bibitem{} Collinge M.J. \et \ 2001, ApJ, 557, 2
\bibitem{} Done C., Mulchaey J.S., Mushotzky R.F., Arnaud K.A., \ 1992, ApJ, 395, 275
\bibitem{} Fabian A.C., Rees M.J., Stella L., White N.E., 1989, MNRAS, 238, 729
\bibitem{} Ferrarese L., Merritt D., \ 2000, ApJ, 539, L9
\bibitem{} Gebhardt K., \et \ 2000, ApJ, 539, L13 
\bibitem{} Haring N., Rix H-W., \ 2004, ApJ, 604, L89 
\bibitem{} Hurkett C.P., \et \ 2008, ApJ, 679, 587 
\bibitem{} Kallman T.R., Palmeri P., Bautista M.A., Mendoza C., Krolik K.H.,  \ 2004, ApJS, 55, 675
\bibitem{} King A.R., \ 2003, ApJ, 596, L27
\bibitem{} King A.R., \ 2005, ApJ, 635, L121
\bibitem{} King A.R., Pounds K.A. \ 2003, MNRAS, 345, 657
\bibitem{} King A.R., \ 2010, MNRAS, 402, 1516
\bibitem{} King A.R,, \ 2010a, MNRAS, 408, 95 
\bibitem{} Lobban A.P., Reeves J.N., Miller L., Turner T.J.,  Braito V., Kraemer S.B., Crenshaw D.M., \ 2011, MNRAS, 414, 1965
\bibitem{} Magdziarz P., Zdziarski A.A., \ 1995, MNRAS, 273, 837
\bibitem{} Matt G., \ 2002, MNRAS, 337, 147
\bibitem{} Pounds K.A, Reeves J.N., King A.R, Page K.L, O'Brien P.T., Turner M.J.L., \ 2003, MNRAS, 345, 705
\bibitem{} Pounds K.A., Reeves J.N., Page K.L., O'Brien P.T., \ 2004, ApJ, 616, 696 
\bibitem{} Pounds K.A., Page K.L., \ 2006, MNRAS, 372, 1275 
\bibitem{} Pounds K.A., Reeves J.N., \ 2009, MNRAS, 397, 249 
\bibitem{} Pounds K.A., Vaughan S., \  2011, MNRAS, 413, 1251 (Paper I)
\bibitem{} Pounds K.A., Vaughan S., \  2011a, MNRAS, 415, 2379 (Paper II)
\bibitem{} Reeves J.N., O'Brien P., Ward M.J., \ 2003, ApJ, 593, L65
\bibitem{} Steenbrugge K.C, Fenovcik M., Kaastra J.S., Costantini E., Verbunt F. \ 2009, A\&A, 496, 107
\bibitem{} Str\"{u}der L., \et  \ 2001, A\&A, 365, L18
\bibitem{} Tombesi F., Cappi M., Reeves J.N., Palumbo G.G.C., Yaqoob T., Braito V., Dadina M., \ 2010, A\&A, 521, 57
\bibitem{} Turner M.J.L., \et  \ 2001, A\&A, 365, L27
\bibitem{} Turner T.J., Miller L., \ 2010, ApJ, 709, 1230 
\bibitem{} Vaughan S., Uttley P., Pounds K.A., Nandra K., Strohmeyer T.E., \ 2011, MNRAS, 413, 2489

\end{thebibliography}
\end{document}